\documentclass{book}
\usepackage{amsmath}
\usepackage{amstext}
\usepackage{amssymb}
\usepackage{mathrsfs}
\usepackage{science}
\usepackage{color}
\usepackage{subfigure}
\usepackage{enumerate}
\usepackage[normalem]{ulem}
\usepackage{graphicx}

\def\ben{\begin{equation}}
\def\een{\end{equation}}
\def\bena{\begin{eqnarray}}
\def\eena{\end{eqnarray}}

\newcommand{\R}{{\mathcal R}}

\newcommand{\edit}[1]{#1}

\newcommand{\ind}[1]{#1}
\newcommand{\subind}[2]{#1}

\begin{document}


\setcounter{chapter}{18}
\chapter{Conserved Charges in Asymptotically (Locally) AdS Spacetimes}

\vspace{-0.5cm}

\begin{minipage}{\textwidth}
\begin{center}
Sebastian Fischetti$^\dagger$, William Kelly$^\dagger$, and Donald Marolf$^{\dagger,\ddagger}$ \\
E-mail: \texttt{sfischet@physics.ucsb.edu}, \texttt{wkelly@physics.ucsb.edu}, \texttt{marolf@physics.ucsb.edu} \\
\vspace{0.25cm}
$^\dagger$\textit{Department of Physics \\ University of California,
Santa Barbara \\ Santa Barbara, CA 93106, USA} \\
\vspace{0.25cm}
$^\ddagger$\textit{Department of Physics \\ University of Colorado, Boulder \\ Boulder, CO 80309, USA}
\end{center}
\end{minipage}
\\

\section{Introduction}
\label{intro}

When a physical system is complicated and non-linear, global symmetries and the associated conserved quantities provide some of the most powerful analytic tools to understand its behavior.  This is as true in theories with a dynamical spacetime metric as for systems defined on a fixed spacetime background.  Chapter 17 has already discussed the so-called Arnowitt-Deser-Misner (ADM) conserved quantities for asymptotically flat dynamical spacetimes, exploring in detail certain subtleties related to diffeomorphism invariance.   In particular, it showed that the correct notion of global symmetry is given by the so-called asymptotic symmetries; equivalence classes of diffeomorphisms with the same asymptotic behavior at infinity.  It was also noted that the notion of asymptotic symmetry depends critically on the choice of boundary conditions. Indeed, it is the imposition of boundary conditions that cause the true gauge symmetries to be only a subset of the full diffeomorphism group and thus allow the existence of non-trivial asymptotic symmetries at all.

This chapter will explore the asymptotic symmetries and corresponding conserved charges of asymptotically anti-de Sitter (AdS) spacetimes (and of the more general asymptotically locally AdS spacetimes).  There are three excellent reasons for doing so.  The first is simply to gain further insight into asymptotic charges in gravity by investigating a new example. Since empty AdS space is a maximally symmetric solution, asymptotically AdS spacetimes are a natural and simple choice.  The second is that the structure one finds in the AdS context is actually much richer than that in asymptotically flat space. At the physical level, this point is deeply connected to the fact (see e.g. \cite{Horowitz:2000fm}) that all multipole moments of a given field in AdS space decay at the same rate at infinity.  So while in asymptotically flat space the far field is dominated mostly by monopole terms (with only sub-leading corrections from dipoles and higher multipoles) all terms contribute equally in AdS.  It is therefore useful to describe not just global charges (e.g., the total energy) but also the local densities of these charges along the AdS boundary.  In fact, it is natural to discuss an entire so-called {\it boundary stress tensor} $T_\mathrm{bndy}^{ij}$ rather than just the conserved charges it defines.  For this reason, we take a somewhat different path to the construction of conserved AdS charges than was followed in chapter 17.  In particular, we will use covariant as opposed to Hamiltonian methods below, though we will show in section \ref{Hamlink} that the end results for conserved charges are equivalent.

The third reason to study conserved charges in AdS is their fundamental relation to the anti-de Sitter/Conformal Field Theory (AdS/CFT) correspondence \cite{Maldacena:1997re,Gubser:1998bc,Witten:1998qj},  which may well be the most common application of general relativity in 21st century physics.  While this is not the place for a detailed treatment of either string theory or AdS/CFT, no Handbook of Spacetime would be complete without presenting at least a brief overview of the correspondence.  It turns out that this is easy to do once we have become familiar with $T_\mathrm{bndy}^{ij}$ and its cousins associated with other (non-metric) fields.  So at the end of this chapter (section \ref{adscft}) we take the opportunity to do so.  We will introduce AdS/CFT from the gravity side without using tools from either string theory or conformal field theory.

We will focus on such modern applications below, along with open questions.  We make no effort to be either comprehensive or historical.  Nevertheless, the reader should be aware that conserved charges for asymptotically AdS spacetimes were first constructed in \cite{Abbott:1981ff}, where the associated energy was also argued to be positive definite.

The plan for this chapter is as follows.  After defining and discussing AdS asymptotics in section \ref{AsympAdSSection}, we construct variational principles for asymptotically AdS spacetimes in section \ref{sec:VPQ}.  This allows us to introduce the boundary stress tensor $T^{ij}_\mathrm{bndy}$ and a similar so-called response function $\Phi_\mathrm{bndy}$ for a bulk scalar field.  The conserved charges $Q[\xi]$ constructed from $T^{ij}_\mathrm{bndy}$ are discussed in section \ref{sec:Qtime} and we comment briefly on positivity of the energy in section \ref{sec:pos}.

Section \ref{Hamlink} then provides a general proof that the $Q[\xi]$ do indeed generate canonical transformations corresponding to the desired asymptotic symmetries.
 As a result, they agree (up to a possible choice of zero-point) with corresponding ADM-like charges $H[\xi]$ that would be constructed via the AdS-analogues of the Hamiltonian techniques used in chapter 17.   The interested reader can find such a Hamiltonian treatment in \cite{Henneaux:1984xu,Henneaux:1985tv,Brown:1986nw}.  Below, we generally consider AdS gravity coupled to a simple scalar matter field.  More complete treatments allowing more general matter fields can be found in e.g. \cite{Papadimitriou:2005ii,Hollands:2005ya,Andrade:2011dg}.  Section \ref{adscft} then defines the algebra ${\cal A}_\mathrm{bndy}$ of boundary observables and provides the above-mentioned brief introduction to AdS/CFT.

\section{Asymptotically Locally AdS Spacetimes} \label{AsympAdSSection}

This section discusses the notion of asymptotically locally AdS spacetimes.
We begin by introducing empty Anti-de Sitter space itself in section \ref{empty} as a maximally-symmetric solution to the Einstein equations.  We then explore the \ind{asymptotic structure} of AdS, and in particular its conformal boundary.  This structure is used to define the notions of asymptotically AdS (AAdS) and asymptotically locally AdS (AlAdS) spacetimes in section \ref{AlAdS}.  Section \ref{sec:FG} then discusses the associated Fefferman-Graham expansion which provides an even more detailed description of the asymptotics and which will play a critical role in constructing variational principles, the boundary stress tensor, and so forth in the rest of this chapter.  Finally, section \ref{sec:diffeos} describes how the above structures transform under diffeomorphisms and introduces the notion of an asymptotic Killing vector field.

\subsection{\ind {Anti-de Sitter Space}}
\label{empty}

Let us begin with a simple geometric description of $(d+1)$-dimensional anti-de Sitter space (AdS${}_{d+1}$) building on the reader's natural intuition for flat geometries.  We will, however, need to begin with a flat spacetime $\mathbb{M}^{2,d}$ of signature $(2,d)$ having two time-directions and $d$ spatial directions, so that in natural coordinates $T^1,T^2,X^1,\dots,X^d$ the line element takes the form

\begin{equation}
\label{M2d}
ds^2 = - (dT^1)^2 -  (dT^2)^2 + (dX^1)^2 + \dots + (dX^{d})^2.
\end{equation}

\begin{figure}
\centering
\includegraphics[width=0.4\textwidth]{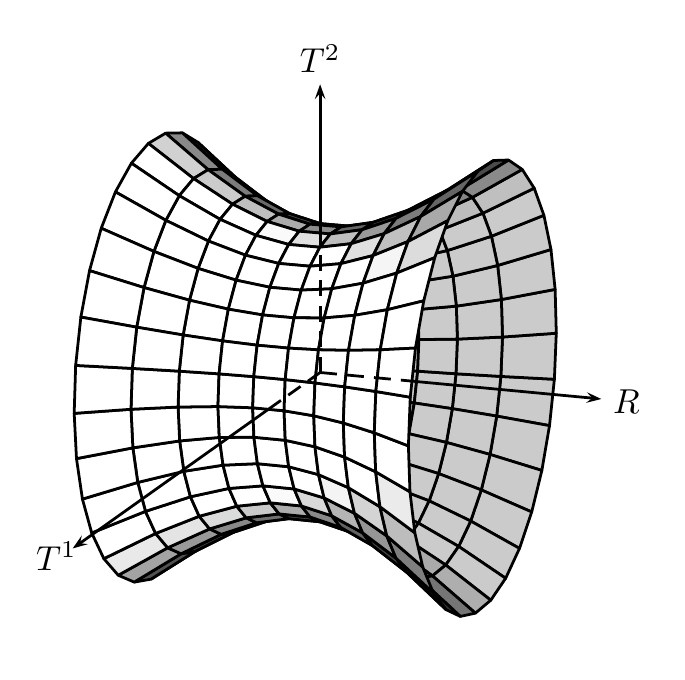}
\caption{The hyperboloid~\eqref{eq:AdShyperboloid} embedded in~$\mathbb{M}^{2,d}$, defining anti-de Sitter space.}
\label{fig:hyperboloid}
\end{figure}

Consider the~$(d+1)$-dimensional hyperboloid ${\cal H}$ of events in $\mathbb{M}^{2,d}$ satisfying
\begin{equation}
\label{eq:AdShyperboloid}
(T^1)^2 + (T^2)^2 - \sum_{i = 1}^{d} \left(X^i\right)^2 = \ell^2,
\end{equation}
and thus which lie at a proper distance $\ell$ from the origin; see figure~\ref{fig:hyperboloid}. This hyperboloid is sometimes known as the $d+1$ anti-de Sitter space AdS${}_{d+1}$, though we will follow a more modern tradition and save this name for a closely related (but much improved!) spacetime that we have yet to introduce.

The isometries of ${\cal H}$ are given by symmetries of~$\mathbb{M}^{2,d}$ preserved by~\eqref{eq:AdShyperboloid}.  Such isometries form the group $SO(d,2)$, generated by the rotation in the $T^1,T^2$ plane together with two copies of the Lorentz group $SO(d,1)$ that act separately on $T^1,X^1,\dots, X^d$ and $T^2, X^1, \dots X^d$.  This gives $(d+1)(d+2)/2$ independent symmetries so that ${\cal H}$ is maximally symmetric.

A simple way to parametrize the hyperboloid is to write~$T^1 = \sqrt{\ell^2+R^2} \, \cos(\tau/\ell)$ and~$T^2 = \sqrt{\ell^2+R^2} \, \sin(\tau/\ell)$, with~$R^2 = \sum (X^i)^2$ so that the induced line element on ${\cal H}$ becomes
\begin{equation}
\label{eq:AdSglobal}
ds^2_{\mathrm{AdS}_{d+1}} = -\left(R^2/\ell^2 + 1\right) \, d\tau^2 + \frac{dR^2}{R^2/\ell^2 + 1} + R^2 \, d\Omega_{d-1}^2.
\end{equation}
On ${\cal H}$, the coordinate~$\tau$ is periodic with period $2\pi$.  But this makes manifest that ${\cal H}$ contains closed timelike curves such as, for example, the worldline $R=0$. It is thus useful to unwrap this time direction by passing to the universal covering space of ${\cal H}$ or, more concretely, by removing the periodic identification of $\tau$ (so that $\tau$ now lives on ${\mathbb R}$ instead of $S^1$).  We will refer to this covering space as the \ind{anti-de Sitter space} AdS${}_{d+1}$ with scale $\ell$.  Of course, the line element remains that of  \eqref{eq:AdSglobal}. Since any Killing field of ${\cal H}$ lifts readily to the covering space, AdS${}_{d+1}$ remains maximally symmetric with isometry group given by (a covering group of) $SO(d,2)$.

The coordinates used in~\eqref{eq:AdSglobal} are called \subind{global coordinates}{anti-de Sitter space}, since they cover all of AdS.  We can introduce another useful set of coordinates, called Poincar\'e coordinates, by setting~$z = \ell^2/\left(T^1 + X^d\right)$,~$t = \ell T^2/\left(T^1 + X^d\right)$, and~$x^i = \ell X^i/\left(T^1 + X^d\right)$ for~$i = 1,\ldots,d-1$.  The metric then becomes
\begin{equation}
\label{eq:AdSPoincare}
ds_{\mathrm{AdS}_{d+1}}^2 = \frac{\ell^2}{z^2}\left(-dt^2 + \sum_{i = 1}^{d-1} \left(dx^i\right)^2 + dz^2 \right).
\end{equation}
Poincar\'e coordinates take their name from the fact that they make manifest a (lower dimensional) Poincar\'e symmetry associated with the $d$ coordinates $t,x^i$.  As is clear from their definitions, these coordinates cover only the region of AdS where $T^1 + X^d > 0$.  This region is called the the \subind{Poincar\'e patch}{anti-de Sitter space}.  While we will not make significant use of \eqref{eq:AdSPoincare} below, we mention these coordinates here since they arise naturally in many discussions of AdS/CFT which the reader may encounter in the future.

Since AdS is maximally symmetric, its Riemann tensor can be written as an appropriately symmetrized combination of metric tensors:
\begin{equation}
\label{eq:Riemann}
R_{\mu\nu\sigma\lambda} = \frac{1}{d(d+1)} \, R \left(g_{\mu\sigma} g_{\nu\lambda} - g_{\mu\lambda} g_{\nu\sigma}\right).
\end{equation}
A computation shows that the scalar curvature of AdS is~$R = -d(d+1)/\ell^2$, and thus that AdS solves the vacuum Einstein field equations with cosmological constant~$\Lambda = -d(d-1)/2\ell^2$:
\begin{equation}
\label{eq:cosmoEFE}
R_{\mu\nu} - \frac{1}{2} \, Rg_{\mu\nu} + \Lambda g_{\mu\nu} = 0.
\end{equation}
In this sense, AdS is a generalization of flat space to~$\Lambda < 0$.

\subsection{Conformal Structure and Asymptotic Symmetries of AdS}
\label{asympt}

\begin{figure}
\label{fig:AdSconformalstrip}
\centering
\subfigure[]{
\includegraphics[width=0.2\textwidth]{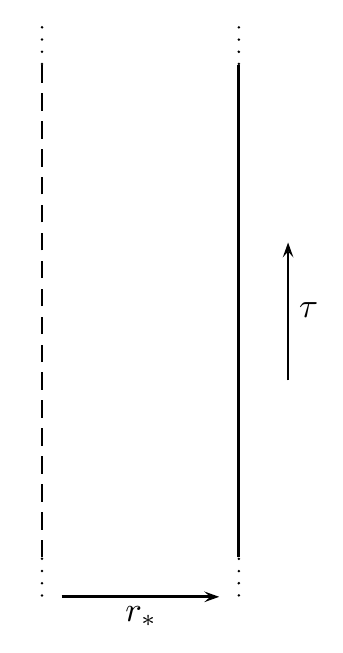}
}
\subfigure[]{
\includegraphics[width=0.18\textwidth]{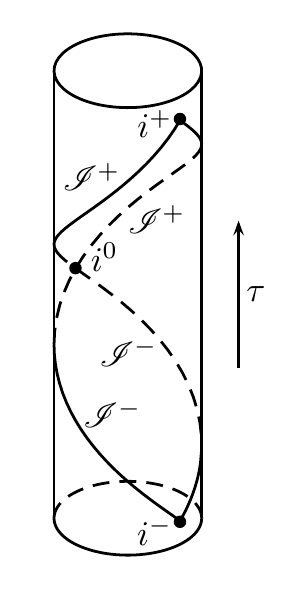}
}
\caption{\subind{Conformal diagrams}{anti-de Sitter space} of AdS$_{d+1}$, showing both the global spacetime and the region covered by the Poincar\'e patch. In both figures, the~$\tau$ direction extends infinitely to the future and to the past.  In~(a), a full~$S^{d-1}$ of symmetry has been suppressed, leaving only the~$\tau$,~$r_*$ coordinates of~\eqref{eq:AdSEinstein}.  The dotted line corresponds to~$r_* = 0$.  In (b), one of the angular directions has been shown explicitly to guide the reader's intuition; the axis of the cylinder corresponds to the dotted line in~(a).  The Poincar\'e patch covers a wedge-shaped region of the interior of the cylinder which meets the boundary at the lines marked $\mathscr{I}^\pm$ and the points marked $i^\pm, i^0$.  These loci form the null, timelike, and spacelike infinities of the associated region (conformal to Minkowski space) on the AdS boundary.}
\end{figure}

We now turn to the asymptotic structure of AdS, which was seen in chapter 17 to be a crucial ingredient in the construction of conserved charges.  It is useful to introduce a new radial coordinate~$r_* = \arctan(R/\ell)$, so that the line element becomes
\begin{equation}
\label{eq:AdSEinstein}
ds^2_{\mathrm{AdS}_{d+1}} = \frac{\ell^2}{\cos^2\left(r_*\right)} \left[ -d\tau^2/\ell^2 + dr_*^2 + \sin^2\left(r_*\right) d\Omega_{d-1}^2\right].
\end{equation}
We can immediately identify~$r_* = \pi/2$ as a conformal boundary, leading to the conformal diagrams shown in Figure~\ref{fig:AdSconformalstrip}.  (For readers not familiar with such diagrams, Chapter 25 will give a brief introduction.)

It is evident from the conformal diagram that AdS is not globally hyperbolic.  In order to evolve initial data on some spacelike surface $\Sigma$ arbitrarily far forward (or backward) in time, one needs to supply additional information in the form of boundary conditions at the conformal boundary. Such boundary conditions will be discussed in detail in section \ref{sec:VPQ}, where they will play critical roles in our discussion of conserved charged.

Although the line element~\eqref{eq:AdSEinstein} diverges at~$r_* = \pi/2$,  the rescaled metric
\begin{equation}
\label{eq:rescaledgglobal}
\hat{g} = \frac{\cos^2(r_*)}{\ell^2} \, g_{\mathrm{AdS}_{d+1}}
\end{equation}
defines a smooth manifold with boundary.  In particular, the metric induced by $\hat{g}$ at~$r_* = \pi/2$ is just that of the flat cylinder~$\mathbb{R} \times S^{d-1}$, also known as the Einstein static universe (ESU). The manifold with boundary will be called $M$ and the boundary itself (at $r_* = \pi/2$) will be called $\partial M$.    Of course, we could equally well have considered the more general rescaled metric
\begin{equation}
\label{eq:rescaledglobalprime}
\hat{g}' = \frac{\cos^2(r_*)}{\ell^2} \, e^{2\sigma} \, g_{\mathrm{AdS}_{d+1}},
\end{equation}
where~$\sigma$ is an arbitrary smooth function on $M$.  This metric is also nonsingular at~$r_* = \pi/2$, but the induced geometry on $\partial M$ is now only conformal to $\mathbb{R} \times S^{d-1}$.
The choice of a particular rescaled metric \eqref{eq:rescaledglobalprime} (or, equivalently, of a particular rescaling factor $\frac{\cos^2(r_*)}{\ell^2} \, e^{2\sigma}$) determines a representative of the corresponding conformal class of boundary metrics.  This choice (which still allows great freedom to choose $\sigma$ away from $\partial M$) is known as the choice of \ind{conformal frame}.  We shall often call this representative ``the boundary metric,'' where it is understood that the above choices must be made for this term to be well-defined.

Although it is not critical for our discussion below, the reader should be aware of the asymptotic structure of the Poincar\'e patch and how it relates to that of global AdS as discussed above.  From ~\eqref{eq:AdSPoincare} we see that the conformal boundary lies at $z = 0$.  The rescaled metric
\begin{equation}
\label{eq:rescaledgPoincare}
\hat{g} = \frac{z^2}{\ell^2} \, g_{\mathrm{AdS}_{d+1}}
\end{equation}
is regular at~$z = 0$, where the induced metric is just~$d$-dimensional Minkowski space.  Now, it is well known~\cite{waldbook} that Minkowski space~$\mathbb{M}^{1,d-1}$ is conformally equivalent to a patch of the Einstein static universe~$\mathbb{R} \times S^{d-1}$.  We conclude that $z=0$ of the Poincar\'e patch is a diamond-shaped piece of~$\partial M$, as shown at right in Figure~\ref{fig:AdSconformalstrip}.

In the interior of AdS the Poincar\'e patch covers a wedge-shaped region.  This can be thought of as follows: future-directed null geodesics fired from~$i^-$ in Figure~\ref{fig:AdSconformalstrip} are focused onto~$i^0$; these geodesics are generators of a null hypersurface which we shall call the past Poincar\'e horizon~$\mathcal{H}^-_\mathrm{Poincar\acute{e}}$.  Likewise, future-directed null geodesics fired from~$i^0$ are focused onto~$i^+$, generating the future Poincar\'e horizon~$\mathcal{H}^+_\mathrm{Poincar\acute{e}}$.  The Poincar\'e patch of AdS is the wedge enclosed by these horizons.

\subsection{A definition of Asymptotically Locally AdS Spacetimes}
\label{AlAdS}

As we saw in chapter 17, when the spacetime metric is dynamical the choice of boundary conditions plays an especially key role in constructions of conserved charges.  In this chapter we consider \ind{boundary conditions} which force the spacetime to behave asymptotically in a manner at least locally similar to \eqref{eq:AdSglobal}.  It turns out to be useful to proceed by using the notion of a conformally rescaled metric $\hat g$ which extends sufficiently smoothly to the boundary (see chapter 25 for further discussion of this technique).  After imposing the equations of motion, this $\hat g$ will allow us to very quickly define both asymptotically AdS (AAdS) and asymptotically local AdS spacetimes (AlAdS).
Below, we follow \cite{Ashtekar:1984zz,Ashtekar:1999jx,Papadimitriou:2005ii,Cheng:2005wk,Skenderis:2000in,Skenderis:2002wp,deHaro:2000xn}.

To begin, recall that our discussion of pure AdS above made use of the fact that the unphysical metrics defined in~\eqref{eq:rescaledgglobal} and~\eqref{eq:rescaledgPoincare} could be extended to the conformal boundary~$\partial M$ of AdS.  We can generalize this notion by considering any manifold $M$ (often called `the bulk') with boundary $\partial M$ and allowing metrics $g$ which are singular on $\partial M$ but for which
but there exists a smooth function~$\Omega$ satisfying~$\Omega|_{\partial M} = 0$,~$(d\Omega)|_{\partial M} \neq 0$ (where $|_{\partial M}$ denotes the pull-back to $\partial M$), and~$\Omega > 0$ on all of $M$, such that
\begin{equation}
\hat{g} = \Omega^2 g
\end{equation}
can be extended to all of $M$ as a sufficiently smooth non-degenerate metric for which the induced metric on $\partial M$ has Lorentz signature.   We will discuss what is meant by sufficiently smooth in more detail in section \ref{sec:FG}, but for the purposes of this section one may take $\hat g$ to be $C^2$ (so that its Riemann tensor is well-defined).  Note that~$\hat{g}$ is not unique; given any allowed~$\Omega$ one is always free to choose
\begin{equation}
\label{eq:Omegap}
\Omega' = e^\sigma \Omega,
\end{equation}
for arbitrary smooth~$\sigma$ on $M$.  Thus, as before, the notion of a particular boundary metric on $\partial M$ is well-defined only after one has chosen some conformal frame.  However, the bulk metric $g$ does induce a unique conformal structure on $\partial M$.  The function~$\Omega$ is termed the \subind{defining function}{conformal frame} of the conformal frame. The above structure is essentially that of Penrose's conformal compactifications \cite{PenroseRindler}, except that the Lorentz signature of $\partial M$ forbids $M$ from being fully compact.  In particular, future and past infinity are not part of $\partial M$.

In vacuum Einstein-Hilbert gravity with cosmological constant \eqref{eq:cosmoEFE}, we define an \ind{asymptotically locally AdS spacetime} to be a spacetime $(g, M)$ as above that solves the Einstein equations \eqref{eq:cosmoEFE}. A key feature of this definition is that it makes no restriction on the conformal structure, or even the topology of the boundary, save that it be compatible with having a Lorentz signature metric.  For an asymptotically locally AdS spacetime to be what we will call \ind{asymptotically AdS}, the induced boundary metric must be conformal to ${\mathbb R} \times S^{d-1}$.
The reader should be aware that in the literature, the term ``asymptotically AdS'' (AAdS) is sometimes used synonymously with ``asymptotically locally AdS'' (AlAdS).  Here we emphasize the distinction between the two for pedagogical purposes, as only AAdS spacetimes can truly be said to approach global AdS near $\partial M$.

To show that AlAdS spacetimes do in fact approach \eqref{eq:Riemann} requires the use of the Einstein equations.   By writing~$g_{\mu\nu} = \Omega^{-2} \hat{g}_{\mu\nu}$, a straightforward calculation then shows~\cite{Skenderis:2002wp} that near $\partial M$ we have
\begin{equation}
\label{eq:RiemannAlAdS}
R_{\mu\nu\sigma\lambda} = -\left|d\Omega\right|^2_{\hat{g}} \left(g_{\mu\sigma} g_{\nu\lambda} - g_{\nu\sigma} g_{\mu\lambda} \right) + \mathcal{O}\left(\Omega^{-3}\right),
\end{equation}
where
\begin{equation}
\left|d\Omega\right|^2_{\hat{g}} \equiv \hat{g}^{\mu\nu} \partial_\mu \Omega \, \partial_\nu \Omega
\end{equation}
extends smoothly to~$\partial M$.  Note that since~$g$ has a second-order pole at~$\partial M$, the leading-order term in~\eqref{eq:RiemannAlAdS} is of order~$\Omega^{-4}$.  The Einstein field equations then imply that
\begin{equation}
\left|d\Omega\right|^2_{\hat{g}} = \frac{1}{\ell^2} \ \ \ {\rm on} \ \partial M.
\end{equation}
It follows that Riemann tensor~\eqref{eq:RiemannAlAdS} of an AlAdS spacetime near $\partial M$ looks like that of pure AdS \eqref{eq:Riemann}.  Further details of the asymptotic structure (and of the approach to \eqref{eq:AdSglobal} for the AAdS case) are elucidated by the Fefferman-Graham expansion near $\partial M$ to which we now turn.

\subsection{The \ind{Fefferman-Graham Expansion}}
\label{sec:FG}

The term asymptotically (locally) AdS suggests that the spacetime metric $g$ should (locally)  approach \eqref{eq:AdSglobal}, at least with a suitable choice of coordinates.  This is far from manifest in the definitions above.  But it turns out to be a consequence of the Einstein equations.  In fact, these equations imply that the asymptotic structure is described by a so-called Fefferman-Graham expansion \cite{FeffermanGraham}.

The basic idea of this expansion is to first choose a convenient set of coordinates and then to attempt a power-series solution to the Einstein equations.  Since the Einstein equations are second order, this leads to a second-order recursion relation for the coefficients of the power series.  For, say, simple ordinary differential equations, one would expect the free data in the power series to be parametrized by two of the coefficients.  The structure that emerges from the Einstein equations is similar, except for the presence of constraint equations similar to those described in chapter 17.  As we briefly describe below, the constraint equations lead to corresponding constraints on the two otherwise free coefficients.  We continue to consider the vacuum case \eqref{eq:cosmoEFE}.

Let us begin by introducing the so-called \ind{Fefferman-Graham coordinates} on some finite neighborhood $U$ of $\partial M$. To do so, note that since the defining function~$\Omega$ is not unique it is possible to choose a~$\sigma$ in~\eqref{eq:Omegap} such that the modified defining function~$z: = \Omega'$ obeys
\begin{equation}
\left|dz\right|^2_{\hat{g}} = \frac{1}{\ell^2}
\end{equation}
on $U$, where $\hat g = z^2 g$.  In fact, we can do so with $\sigma|_{\partial M} =1$ so that we need not change the conformal frame.  We can then take the defining function~$z$ to be a coordinate near the boundary; the notation $z$ is standard for this so-called ``Fefferman-Graham radial coordinate.''  We choose the other coordinates $x^i$ to be orthogonal to $z$ in $U$ (according to the metric $\hat g$).  The metric in these so-called Fefferman-Graham coordinates will then take the form
\begin{equation}
\label{eq:FG}
ds^2 = \frac{\ell^2}{z^2}\left(dz^2 + \gamma_{ij}(x,z) \, dx^i \, dx^j\right),
\end{equation}
where~$i = 0,\ldots,d$.  By construction,~$\gamma_{ij}$ can be extended to~$\partial M$, so it should admit an expansion (at least to some order) in non-negative powers of~$z$:
\begin{equation}
\label{eq:intFG}
\gamma_{ij}(x,z) = \gamma^{(0)}_{ij}(x) + z \gamma^{(1)}_{ij}(x) + \cdots.
\end{equation}
Note that $\gamma^{(0)}_{ij}$ defines the metric $\gamma^{(0)}$ on $\partial M$ in this conformal frame.

Since the Einstein equations are second order partial differential equations, plugging in the ansatz \eqref{eq:intFG} leads to a second order recursion relation for the $\gamma^{(n)}$.  For odd $d$ this recursion relation admits solutions for all $\gamma^{(n)}$.  After specifying $\gamma^{(0)}$, one finds that all $\gamma^{(n)}$ with $n < d$ are uniquely determined (and, in fact $\gamma^{(n)}$ vanishes for all odd $n < d$).  For example, for $d > 2$ one finds~\cite{deHaro:2000xn}\footnote{\edit{We caution the reader to be wary of the differing sign conventions in the literature.  For example, the sign conventions for Riemann and extrinsic curvatures used in~\cite{deHaro:2000xn} are opposite from the ones used here.}}
\begin{equation}
\label{eq:g2}
\gamma^{(2)}_{ij} =  \edit{-}\frac{1}{d-2} \left( \R_{ij} - \frac{1}{2(d-1)} \R \gamma^{(0)}_{ij} \right),
\end{equation}
where ${\cal R}, {\cal R}_{ij}$ are respectively the Ricci tensor and Ricci scalar of $\gamma^{(0)}$.

However, new data enters in $\gamma^{(d)}$.  This new data is subject to constraints analogous to those discussed in the Hamiltonian formalism in chapter 17.  Indeed, these constraints may be derived by considering the analogues of the Hamiltonian and momentum constraints on surfaces with $z=constant$.  They determine the trace and divergence of $\gamma^{(d)}$ (again for $d$ odd) through
\begin{equation}
\label{dodd}
\left( \gamma^{(0)} \right)^{ij} \gamma^{(d)}_{ij} =0, \hspace{1cm} \left( \gamma^{(0)} \right)^{ki} D_k \gamma^{(d)}_{ij} =0,
\end{equation}
where $D_k$ is the $\gamma^{(0)}$-compatible derivative operator on $\partial M$ (where we think of all $\gamma^{(n)}$ as being defined).  We will give a short argument for \eqref{dodd} in section \ref{sec:Qtime}.  Once we have chosen any $\gamma^{(d)}$ satisfying \eqref{dodd}, the recursion relation can then be solved order-by-order to express all higher $\gamma^{(n)}$ in terms of $\gamma^{(0)}$ and $\gamma^{(d)}$.  Of course, the series \eqref{eq:FG} describes only the asymptotic form of the metric.  There is no guarantee that there is in fact a smooth solution in the interior matching this asymptotic data, or that such a smooth interior solution is unique when it exists.

The situation is slightly more complicated for even $d$, where the recursion relations for the ansatz \eqref{eq:intFG} break down at the order at which~$\gamma^{(d)}$ would appear.  To proceed,  one must allow logarithmic terms to arise at this order and use the more general ansatz
\begin{equation}
\label{eq:FGexpansion}
\gamma_{ij}(x,z) = \gamma^{(0)}_{ij} + z^2 \gamma^{(2)}_{ij} + \cdots + z^d \gamma^{(d)}_{ij} + z^d \bar{\gamma}^{(d)}_{ij} \log \, z^2 + \cdots,
\end{equation}
where, since the structure is identical for all $d$ up to order $n=d$, we have made manifest that $\gamma^{(n)}=0$ for all odd $n < d$.  The higher order terms represented by $\cdots$ include both higher even powers of $z$ and such terms multiplied by $\log z$.  One finds that $\bar \gamma ^{(d)}$ is fully determined by $\gamma^{(0)}$ and satisfies
\begin{equation}
\label{eq:gbd}
\left(\gamma^{(0)}\right)^{ij} \bar{\gamma}^{(d)}_{ij} = 0, \hspace{1cm} \left( \gamma^{(0)} \right)^{ki} D_k  \bar{\gamma}^{(d)}_{ij} = 0.
\end{equation}
For example, for $d=2$,~$4$, one obtains~\cite{deHaro:2000xn}
\begin{align}
\label{eq:gb2}
\bar \gamma^{(2)}_{ij} &= 0, \\
\label{eq:gb4}
\bar \gamma^{(4)}_{ij} &= \frac{1}{8} \, \R_{ikjl} \R^{kl} \edit{-} \frac{1}{48} \, D_i D_j \R \edit{+} \frac{1}{16} \, D^2 \R_{ij} - \frac{1}{24} \, \R \R_{ij} + \left(\edit{-}\frac{1}{96} \, D^2 \R + \frac{1}{96} \, \R^2 - \frac{1}{32} \, \R_{kl} \R^{kl}\right) \gamma^{(0)}_{ij},
\end{align}
where~$\R_{ijkl}$ is the Riemann tensor of~$\gamma^{(0)}$, and indices are raised and lowered with~$\gamma^{(0)}$.  But $\gamma^{(d)}$ may again be chosen freely subject to dimension-dependent conditions that fix its divergence and trace.  As examples, one finds~\cite{deHaro:2000xn}
\begin{align}
d=2: & \left(\gamma^{(0)}\right)^{ij} \gamma^{(d)}_{ij} = \edit{-}\frac{1}{2} \, {\cal R}, \hspace{1cm} D^i  \gamma^{(d)}_{ij} = \edit{-}\frac{1}{2} D_j {\cal R}, \\
d=4: & \left(\gamma^{(0)}\right)^{ij} \gamma^{(d)}_{ij} = \frac{1}{16}\left(\R_{ij} \R^{ij} - \frac{2}{9} \, \R^2\right),
	\\ & D^i \gamma^{(d)}_{ij} = \frac{1}{8} {{\cal R}_i}^k D^i {\cal R}_{kj} - \frac{1}{32} D_j \left({\cal R}^{ik} {\cal R}_{ik}\right) + \frac{1}{288} {\cal R} D_j {\cal R}.
\label{traces}
\end{align}
The higher terms in the series are again uniquely determined by $\gamma^{(0)}$, $\gamma^{(d)}$.

In general, the terms $\gamma^{(n)}$ become more and more complicated at each order.  But the expansion simplifies when $\gamma^{(0)}_{ij}$ is conformally flat and~$\gamma^{(d)}_{ij} = 0$.  In this case one finds~\cite{Skenderis:1999nb} that the recursion relation can be solved exactly and terminates at order~$z^4$.  In particular, the bulk metric so obtained is also conformally flat, and is thus locally AdS$_{d+1}$.  For $d = 2$, the Fefferman-Graham expansion can be integrated exactly for any $\gamma^{(0)}$,~$\gamma^{(d)}$, and always terminates at order~$z^4$ to define a metric that is locally AdS${}_3$.

\subsection{Diffeomorphisms and symmetries in AlAdS}
\label{sec:diffeos}

The reader of this Handbook is by now well aware of the important roles played by diffeomorphisms in understanding gravitational physics.  Let us therefore pause briefly to understand how such transformations affect the structures defined thus far.  We are interested in diffeomorphisms of our manifold $M$ with boundary $\partial M$.  By definition, any such diffeomorphism must map $\partial M$ to itself; i.e., it also induces a diffeomorphism of $\partial M$.  As usual in physics, we consider diffeomorphisms (of $M$) generated by vector fields $\xi$; the corresponding diffeomorphism of $\partial M$ is generated by some $\hat \xi$, which is just the restriction of $\xi$ to $\partial M$ (where by the above it must be tangent to $\partial M$).

Of course, the metric $g$ transforms as a tensor under this diffeomorphism.  But if we think of the diffeomorphism as acting only on dynamical variables of the theory then the defining function $z= \Omega$ does not transform at all, and in particular does not transform like a scalar field.  This means that the rescaled metric $\hat g = z^2 g$ does {\it not} transform like a tensor, and neither does the boundary metric $\gamma^{(0)}$.  Instead, the diffeomorphism induces an additional conformal transformation on $\partial M$; i.e., a change of conformal frame.

We can make this explicit by considering \subind{diffeomorphisms}{asymptotically locally AdS} that preserve the Fefferman-Graham gauge conditions; i.e., which satisfy
\begin{equation}
\label{eq:preserveFG}
\delta g_{zz} = 0 = \delta g_{iz}
\end{equation}
for
\begin{equation}
\delta g_{\mu\nu} = \pounds_\xi g_{\mu\nu}
=\nabla_{\mu} \xi_{\nu} + \nabla_{\nu} \xi_{\mu},
\end{equation}
where we use $\pounds_\xi$ to denote Lie derivatives along $\xi$ and $\nabla_\mu$ is the covariant derivative compatible with the metric $g$ on $M$.
Let us decompose the components~$\delta g_{\mu\nu}$ into
\begin{align}
\label{eq:Liezz}
\pounds_\xi g_{zz} &= \frac{2\ell}{z} \, \partial_z \left(\frac{\ell}{z} \, \xi^z \right), \\
\label{eq:Lieiz}
\pounds_\xi g_{iz} &= \frac{\ell^2}{z^2} \left( \partial_i \xi^z  + \gamma_{ij} \partial_z \xi^j \right), \\
\label{eq:Lieij}
\pounds_\xi g_{ij} &= \frac{\ell^2}{z^2} \left({\pounds}_{\hat{\xi}} \gamma_{ij} +z^2 \, \partial_z \left(z^{-2} \gamma_{ij} \right) \xi^z \right),
\end{align}
where~${\pounds}_{\hat{\xi}}$ is the Lie derivative  with respect to~$\hat{\xi}$ on $\partial M$.  These conditions can be integrated using \eqref{eq:preserveFG} to obtain
\begin{align}
\label{eq:xiz}
\xi^z &= z \hat{\xi}^z(x), \\
\label{eq:xii}
\xi^i &= \hat{\xi}^i(x) - \partial_j \hat{\xi}^z \, \int_0^z z' \gamma^{ji}(z') \, dz',
\end{align}
where $\hat \xi^z$ and $\hat \xi^i$ are an arbitrary function and vector field on $\partial M$ (which we may transport to any $z=constant$ surface by using the given coordinates to temporarily identify that surface with $\partial M$).  In particular, for $\hat \xi^i = 0$ we find
\begin{equation}
g_{ij} + \delta g_{ij} = \frac{\ell^2}{z^2}\left(1 - 2\hat{\xi}^z\right)\gamma^{(0)}_{ij} + \mathcal{O}(z^0).
\end{equation}
Thus the boundary metric transforms as $\gamma^{(0)} \rightarrow e^{-2\hat{\xi}^z} \gamma^{(0)}_{ij}$.  Such transformations are called \ind{conformal transformations} by relativists and Weyl transformations by particle physicists; we will use the former, but the reader will find both terms in various treatments of AlAdS spacetimes.  This is precisely the change of conformal frame mentioned above.

Let us now turn to the notion of \ind{symmetry}.  As in chapter 17, we might be interested either in an exact symmetry of some metric $g$, generated by a Killing vector field (KVF) satisfying $\nabla_{(\nu} \xi_{\mu)} =0$, or in some notion of asymptotic symmetry.  We will save the precise definition of an asymptotic symmetry for section \ref{sec:CTs} as, strictly speaking, this first requires the construction an appropriate variational principle and a corresponding choice of boundary conditions.  However, we will discuss the closely related (but entirely geometric) notion of an asymptotic Killing field below.

Suppose first that $\xi$ is indeed a KVF of $g$ so that $\pounds_\xi g =0$.  It is clear that there are two cases to consider.  Either $\pounds_\xi \Omega =0$ (in which case we say that $\xi$ is compatible with $\Omega$) or
$\pounds_\xi \Omega \neq 0$ (in which case we say that $\xi$ is not compatible with $\Omega$).  In the former case we clearly have $\pounds_\xi \hat g = \pounds_\xi (\Omega^2 g) =0$ so that $\xi$ is also a Killing field of $\hat g$.  But more generally we have seen that the corresponding diffeomorphism changes $\hat g$ by a conformal factor.  The generators of such diffeomorphisms are called \ind{conformal Killing fields} of $\hat g$ (see e.g. Appendix C.3 of~\cite{waldbook}) and satisfy
\begin{equation}
\label{eq:conformalKilling}
\pounds_\xi \hat{g}_{\mu\nu} = (\pounds_\xi \ln \Omega^2) \hat g_{\mu\nu} \Rightarrow 2\widehat\nabla_{(\mu}\xi_{\nu)} = \frac{2}{d+1} \, \left(\widehat \nabla_\sigma \xi^\sigma\right) \hat{g}_{\mu\nu},
\end{equation}
where $\widehat \nabla$ is the covariant derivative compatible with $\hat g$, and indices on~$\xi^\mu$ are lowered with~$\hat{g}_{\mu\nu}$.  Note that the induced vector field $\hat \xi$ on $\partial M$ is again a conformal Killing field of $\gamma^{(0)}$.

This suggests that we define an \ind{asymptotic Killing field} to be any vector field $\xi$ that satisfies \eqref{eq:conformalKilling} to leading order in $\Omega$ at $\partial M$.  If we ask that $\xi$ also preserve Fefferman-Graham gauge we may then expand~\eqref{eq:xiz} and~\eqref{eq:xii} and insert into~\eqref{eq:conformalKilling} to obtain
\begin{align}
\xi^z &= z \hat{\xi}^z(x), \\
\xi^i &= \hat{\xi}^i(x) - \frac{1}{2} \, z^2 \left(\gamma^{(0)}\right)^{ij} \partial_j \hat{\xi}^z + \mathcal{O}(z^4),
\label{eq:asKVF}
\end{align}
\begin{equation}
\label{eq:conformalxi}
{\pounds}_{\hat{\xi}} \gamma_{ij}^{(0)} - \frac{2}{d+1} \, \left(D_k \hat{\xi}^k + \hat{\xi}^z\right) \gamma^{(0)}_{ij} = 0.
\end{equation}
Taking the trace of the condition \eqref{eq:conformalxi} shows that $\hat{\xi}^z = \frac{1}{d}D_i \hat{\xi}^i$, so~\eqref{eq:conformalxi} is the conformal Killing equation for~$\hat{\xi}$ with respect to~$\gamma^{(0)}$. In other words, conformal Killing fields $\hat \xi$ of $\gamma^{(0)}$ are in one-to-one correspondence with asymptotic Killing fields of $g$ which preserve Fefferman-Graham gauge, where the equivalence relation is given by agreement to the order shown in~\eqref{eq:asKVF}.

\subsection{Gravity with \ind{Matter}}
\label{sec:GravwithMatter}

Our treatment above has focused on vacuum gravity.  It is useful to generalize the discussion to include matter fields, both to see how this influences the above result and also to better elucidate the general structure of asymptotically AdS field theory. Indeed, readers new to dynamics in AdS space will gain further insight from section \ref{sec:FG} if they re-read it after studying the treatment of the free \ind{scalar field} below. We use a single scalar as an illustrative example of matter fields; see  \cite{Papadimitriou:2005ii,Hollands:2005ya} for more general discussions.

For simplicity, we first consider a massive scalar field in a fixed AlAdS$_{d+1}$ gravitational background, which we take to be in Fefferman-Graham form~\eqref{eq:FG}.  This set-up is often called the \ind{probe approximation} as it neglects the back-reaction of the matter on the spacetime.  The action is as usual
\begin{equation}
\label{eq:scalarS}
S_{\phi}^{Bulk} = -\frac{1}{2}\int d^{d+1}x \, \sqrt{|g|} \, \left(g^{\mu\nu} \partial_\mu \phi \partial_\nu \phi + m^2 \phi^2 \right).
\end{equation}
We study the behavior of solutions near the boundary~$z = 0$ by seeking solutions which behave at leading order like~$z^\Delta$ for some power~$\Delta$.  The equation of motion
\begin{equation}
\label{eq:scalarEOM}
\left(-\Box + m^2\right)\phi = 0
\end{equation}
then requires~$(m\ell)^2 = \Delta(\Delta - d)$, yielding two independent small-$z$ behaviors~$z^{\Delta_\pm}$.  Here we have defined~$\Delta_\pm = d/2 \pm \nu$, with~$\nu \equiv \sqrt{(d/2)^2 + (m\ell)^2}$.  A priori, it seems that we should consider only~$\nu \geq \nu_\mathrm{min}$ for some~$\nu_\mathrm{min} > 0$, since one might expect~$(m\ell)^2 \geq 0$.  However, it can be shown~\cite{Breitenlohner1982} that scalar fields with small tachyonic masses in AdS$_{d+1}$ are stable as long as the mass satisfies the so-called \ind{Breitenlohner-Freedman (BF) bound}~$(m\ell)^2 \geq  -d^2/4 =: m^2_{BF}$; we therefore consider~$\nu \geq 0$.  The essential points here are: i) It is only for $|(m\ell)^2| \gg 1$ that the flat-space approximation must hold, so for small $|(m\ell)^2|$ the behavior can differ significantly from that of flat space; ii) as noted above, the fact that AdS is not globally hyperbolic means that we must impose boundary conditions at $\partial M$.  These boundary conditions generally require $\phi$ to vanish on $\partial M$.  So even for $m^2=0$ we would exclude the `zero mode' $\phi = \mathrm{constant}$.  For a given boundary condition, the spectrum of modes turns out to be discrete.  As a result, we may lower $m^2$ a finite amount below zero before a true instability develops.

The asymptotic analysis above suggests that we seek a solution of the form
\begin{equation}
\label{eq:scalarexpansionGen}
\phi(x,z) = z^{\Delta_-} \left(\phi^{(0)} + z^2 \phi^{(2)} + \cdots \right) + z^{\Delta_+} \left(\phi^{(2\nu)} +  z^2 \phi^{(2\nu + 2)}  + \cdots \right).
\end{equation}
For non-integer $\nu$ the equation of motion can be solved order-by-order in~$z$ to uniquely express all coefficients in terms of $\phi^{(0)}$ and $\phi^{(2\nu)}$.  But for integer $\nu$ the difference $\Delta_+ - \Delta_-$ is an even integer and the two sets of terms in \eqref{eq:scalarexpansionGen} overlap.   This notational issue is connected to a physical one:  keeping only even-integer powers of $z$ (times $z^{\Delta_-}$) does not allow enough freedom to solve the resulting recursion relation; there is no solution at order~$d-2\Delta_-$.  To continue further we must introduce a logarithmic term and write:
\begin{equation}
\label{eq:scalarexpansionInt}
\phi(x,z) = z^{\Delta_-} \left(\phi^{(0)} + z^2 \phi^{(2)} + \cdots \right) + z^{\Delta_+} \log z^2 \left(\psi^{(2\nu)} +  z^2 \psi^{(2\nu + 2)}  + \cdots \right).
\end{equation}
The recursion relations then uniquely express all coefficients in terms of the free coefficients~$\phi^{(0)}$ and~$\phi^{(2\nu)}$.  As an example, we note for later purposes that (for any value of $\nu$)
\begin{equation}
\label{eq:p2}
\phi^{(2)} = \frac{1}{4(\nu-1)} \Box^{(0)} \phi^{(0)},
\end{equation}
where $\Box^{(0)}$ is the scalar wave operator defined by $\gamma^{(0)}$ on $\partial M$.  Dimensional analysis shows that the higher coefficients $\phi^{(n)}$ for integer $n < 2 \Delta_+ -d$ involve $n$ derivatives of $\phi^{(0)}$.

We now couple our scalar to dynamical gravity using
\begin{equation}
S = S_\mathrm{grav} + S_{\phi}^\mathrm{Bulk},
\end{equation}
where~$S_\mathrm{grav}$ is the action for gravity.  We will postpone a discussion of boundary terms to section \ref{sec:VPQ}; for now, we simply focus on solving the resulting equations of motion
\begin{equation}
\label{eq:sourcedEFE}
R_{\mu\nu} - \frac{1}{2} \, Rg_{\mu\nu} + \Lambda g_{\mu\nu} = 8\pi G T^\mathrm{(matter)}_{\mu\nu}.
\end{equation}
As in the vacuum case we write the metric in the form~\eqref{eq:FG}, and as in the solution for nondynamical gravity we write the scalar field as in~\eqref{eq:scalarexpansionInt}.  Note that we keep the logarithmic term in~\eqref{eq:FGexpansion} for all $d$ as, depending on the matter content, it may be necessary even for odd~$d$. (When it is not needed, the equations of motion force its coefficient $\bar \gamma_d$ to vanish.)  The stress tensor of the scalar field then behaves like
\begin{equation}
\label{eq:scalarTmunu}
T_{\mu\nu}^{(\mathrm{matter})} dx^\mu dx^\nu = \Delta_- z^{2(\Delta_- - 1)} \left[\frac{d}{2} \left(\phi^{(0)}\right)^2 dz^2 + z\phi^{(0)} \partial_i \phi^{(0)} \, dz \, dx^i + \nu \left(\phi^{(0)}\right)^2 \gamma^{(0)}_{ij} dx^i \, dx^j + \cdots \right].
\end{equation}
For~$\Delta_- < 0$ and $\phi^{(0)} \neq 0$, the matter stress tensor turns out to diverge too rapidly at~$z = 0$ for the equations of motion to admit an AlAdS solution.  So for $\Delta_- < 0$ the only scalar field boundary condition consistent with the desired physics is $\phi^{(0)} =0$. But for $\Delta _- \ge 0$ the equations of motion \textit{do} admit AlAdS solutions with $\phi^{(0)} \neq 0$ and further input is required to determine the boundary conditions.  We will return to this issue in section \ref{sec:scalarS}.

Evidently, the equations of motion admit solutions of the forms~\eqref{eq:FG} and~\eqref{eq:scalarexpansionInt} only if the components of the matter stress tensor in Fefferman-Graham coordinates diverge as~$1/z^2$ or slower.  This result allows us to generalize our definition of asymptotically locally AdS spacetimes to include matter: an \subind{AlAdS spacetime with matter}{asymptotically locally AdS} is a manifold $M$ as above with fields satisfying the equations of motion and the requirement that $\Omega^2 T_{\mu\nu}$ admits a continuous limit to~$\partial M$.

\section{Variational principles and charges} \label{sec:VPQ}

Noether's theorem teaches us that variational principles provide a powerful link between symmetries and conservation laws, allowing the latter to be derived without detailed knowledge of the equations of motion.  This procedure works as well for gravitational theories as for systems defined on a fixed spacetime background, though there is one additional subtlety.  In more familiar theories, it is often sufficient to consider only variations of compact support so that all boundary terms arising from variations of an action can be discarded.  But as shown in chapter 17 in the asymptotically flat context, when the gravitational constraints (which are just certain equations of motion!) are satisfied
the gravitational charges become pure boundary terms with no contributions from the bulk.   Discarding all boundary terms in Noether's theorem would thus lead to trivial charges and we will instead need to treat boundary terms with care.  It is in part for this reason that we refer to \ind{{\it variational principles}} as opposed to mere actions, the distinction being that all variations of the former vanish when the equations of motion and boundary conditions hold, even including any boundary terms that may arise in computing the variations.  Constructing a good variational principle generally requires that we add boundary terms to the familiar bulk action, and that we tailor the choice of such boundary terms to the boundary conditions we wish to impose on $\partial M$.

\subsection{A toy model of AdS: Gravity in a box}
\label{sec:DIB}

\begin{figure}
\centering
\includegraphics[width=0.4\textwidth]{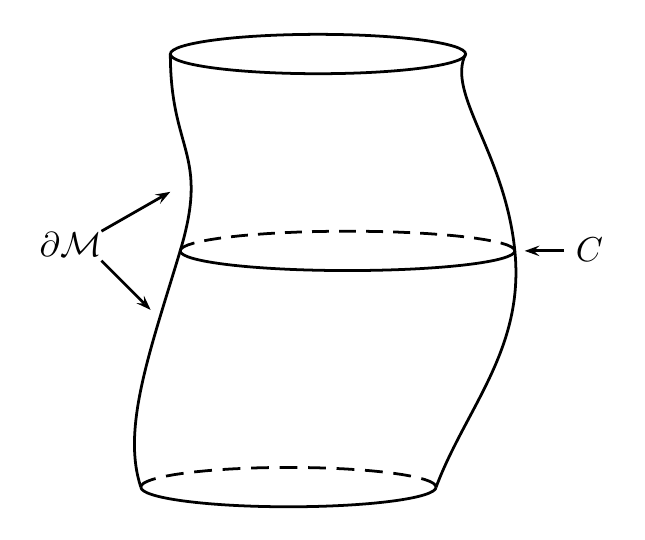}
\caption{A sketch of the spacetime ${\cal M}$.  The codimension two surface $C$ is a Cauchy surface of the boundary $\partial M$.}
\label{fig:Bulkboundary}
\end{figure}

We have seen that AlAdS spacetimes are conformally equivalent to manifolds with timelike boundaries.   This means that (with appropriate boundary conditions) light signals can bounce off of $\partial M$ and return to the interior in finite time,  boundary conditions are needed for time evolution, and indeed much of physics in AlAdS spacetimes is indeed like field theory in a finite-sized box.   This analogy also turns out to hold for the study of conservation laws in theories with dynamical gravity.  It will therefore prove useful to first study conservation laws for gravity on a manifold $M$ with a finite-distance timelike boundary $\partial M$, which will serve as a toy model for AlAdS gravitational dynamics.  This subject, which we call \subind{``gravity in a box''},{Variational Principle} was historically studied for its own sake by Brown and York \cite{Brown:1992br}.  We largely follow their approach below. For simplicity we will assume that $\partial M$ is globally hyperbolic with compact Cauchy surfaces as shown in figure~\ref{fig:Bulkboundary}, though the more general case can typically be treated by imposing appropriate boundary conditions in the asymptotic regions of $\partial M$.

Out first task is to construct a good variational principle.  But as noted above this will generally require us to add boundary-condition-dependent boundary terms to the bulk action.  It is thus useful to have some particular boundary condition (or, at least, a class of such conditions) in mind before we begin.  In scalar field theory, familiar classes of boundary conditions include the Dirichlet condition ($\phi|_{\partial M}$ fixed, so $\delta \phi|_{\partial M} =0$), the Neumann condition (which fixes the normal derivative), or the more general class of Robin conditions (which fix a linear combination of the two).  All of these have analogues for our gravity in a box system, but for simplicity we will begin with a Dirichlet-type condition.  Recall from chapter 18 that, when discussing the initial value problem, the natural initial data on a Cauchy surface consists of the induced metric and the extrinsic curvature (or, equivalently, the conjugate momentum as described in chapter 17).
Since the equations of motion are covariant, the analysis of possible boundary conditions on timelike boundaries turns out to be very similar so that the natural Dirichlet-type condition is to fix the induced metric $h_{ij}$ on $\partial M$.

An important piece of our variational principle will of course be
the \ind{Einstein-Hilbert action} $S_{EH} = \frac{1}{2\kappa}\int \sqrt{-g}  \ R$ (with $\kappa = 8 \pi G$).  But $S_{EH}$ is not sufficient by itself as a standard calculation  gives
\begin{align}
\label{delSeh}
\delta S_{EH} &= \delta \left( \frac{1}{2\kappa} \int_{M} \sqrt{-g} R \right) \cr
&=\frac{1}{2\kappa} \int_{M} \sqrt{-g} \edit{\left( R_{\mu\nu} - \frac{1}{2} R g_{\mu\nu} \right) \delta g^{\mu\nu}} + \frac{1}{2\kappa} \int_{\partial M} \sqrt{|h|} \hat r_\lambda G^{\mu\nu\rho\lambda} \nabla_\rho \delta g_{\mu\nu},
\end{align}
where $\hat r^\lambda$ is the outward pointing unit normal to $\partial M$ and
\begin{align}
G^{\mu\nu\rho\lambda} = g^{\mu(\rho} g^{\lambda)\nu} - g^{\mu\nu} g^{\rho\lambda}.
\end{align}
In \eqref{delSeh} we have discarded boundary terms not associated with $\partial M$ (i.e., boundary terms in any asymptotic regions of $M$) as they will play no role in our analysis.  Nevertheless, the second term in~\eqref{delSeh} (the boundary term) generally fails to vanish for useful boundary conditions, so that $S_{EH}$ is not fully stationary on solutions.

However, when $\delta h_{ij}=0$ this problem term turns out to be an exact variation of another boundary term, known as the \ind{Gibbons-Hawking term}, given by the integral of the trace of the extrinsic curvature of ${\partial M}$.
(For related reasons the addition of this term is necessary when constructing a gravitational path integral, see~\cite{hawking-79}).    As a result, enforcing the boundary condition $\delta h_{ij} =0$ guarantees that all variations of the action
\begin{equation}
\label{eq:DIB}
S_\mathrm{Dirichlet \ in \ a \ box} = S_{EH} + S_{GH} = \frac{1}{2\kappa} \int_{\cal M} \sqrt{-g} R - \frac{1}{\kappa} \int_{\partial M} \sqrt{|h|} K
\end{equation}
\edit{vanish precisely when the bulk equations of motion hold.  Here, $K= h_{ij} K^{ij}$ is the trace of the extrinsic curvature on ${\partial M}$, with~$K_{ij} = - (\pounds_n h_{ij})/2$, where~$n$ is the outward-pointing unit normal to~$\partial M$.} Thus \eqref{eq:DIB} gives a good variational principle for our Dirichlet problem.

Now, \ind{Noether's theorem} teaches us that every continuous symmetry of our system should lead to a conservation law (though the conservation laws associated with pure gauge transformations are trivial).  Gravity in a box is defined by the action \eqref{eq:DIB} and by the choice of some Lorentz-signature metric $h_{ij}$ on $\partial M$.  The first ingredient, the action \eqref{eq:DIB}, is manifestly invariant under any diffeomorphisms of $M$.  Such diffeomorphisms are generated by vector fields $\xi$ on $M$ that are tangent to $\partial M$ at the boundary (so that the diffeomorphism maps $\partial M$ to itself).  As before, we use $\hat \xi$ to denote the induced vector field on $\partial M$.  The associated diffeomorphism of $M$ will preserve $h_{ij}$ if $\hat \xi$ is a Killing field on the boundary. As discussed in chapter 17, a diffeomorphism supported away from the boundary should be pure gauge.  So it is natural to expect that the \ind{asymptotic symmetries} of our system are classified by the choice of boundary Killing field $\hat \xi$, with the particular choice of a bulk extension $\xi$ being pure gauge.

This set up should remind the reader of (non-gravitational) field theories on fixed spacetime backgrounds.  There one finds conservation laws associated with each Killing field of the background metric.  Here again the conservation laws are associated with Killing fields of the background structure, though now the only such structure is the boundary metric $h_{ij}$.

Pursuing this analogy, let us recall the situation for field theory on a fixed (non-dynamical) spacetime background.  There, Noether's theorem for global symmetries (e.g., translations along some Killing field $\xi_{KVF}$) would instruct us to vary the action under a space-time generalization of the symmetry (e.g., diffeomorphism along $f(x) \xi_{KVF}$ for general smooth functions $f(x)$, or more generally under arbitrary diffeomorphisms).   It is clear that the analogue for gravity in a box is just to vary \eqref{eq:DIB} under a general diffeomorphism of $M$.

It turns out to be useful to do so in two steps.  Let us first compute an arbitrary variation of \eqref{eq:DIB}.  By construction, it must reduce to a boundary term when the equations of motion hold, and it must vanish when $\delta h_{ij} =0$.  Thus it must be linear in $\delta h_{ij}$.
A direct calculation (see appendix E of~\cite{waldbook}) gives
\begin{equation}
\label{eq:varyDIB}
\delta S_\mathrm{Dirichlet \ in \ a \ box} = \frac{1}{2} \int_{\partial {M}} \sqrt{|h|}  \tau^{ij} \delta h_{ij},
\end{equation}
where $\tau^{ij} =  \kappa^{-1}(K^{ij} - K h^{ij})$.  This $\tau^{ij}$ is sometimes referred to as the radial conjugate momentum since it has the same form as the (undensitized) conjugate momentum introduced on spacelike surfaces in chapter 17.  This agreement of course follows from general principles of Hamilton-Jacobi theory.  The reader should recall that for field theory in a fixed spacetime background the functional derivative of the action with respect to the metric defines the field theory stress tensor.  By analogy, the object $\tau^{ij}$ defined above is often called the \subind{boundary stress tensor}{stress tensor} (or the \subind{Brown-York stress tensor}{stress tensor}) of the gravitational theory.

Let us now specialize to the case where our variation is a diffeomorphism of $M$. As we have seen, $\xi$ also induces a diffeomorphism of the boundary $\partial M$ generated by some $\hat \xi$.  Then $\delta h_{ij} = D_i \hat \xi_j  +D_j \hat \xi_i$, where $D_i$ is the covariant derivative compatible with $h_{ij}$.  Using the symmetry of $\tau^{ij} = \tau^{ji}$ we find

\begin{equation}
\label{eq:BYcons}
\delta S_\mathrm{Dirichlet \ in \ a \ box} = \int_{\partial {M}} \sqrt{|h|}  \tau^{ij} D_i \hat \xi_j = \edit{-} \int_{\partial {M}} \sqrt{|h|}  \hat \xi_j D_i \tau^{ij} ,
\end{equation}
where in the last step we integrate by parts and take $\hat \xi$ to have compact support on $\partial M$ so that we may discard any boundary terms.  Since $\hat \xi$ is otherwise arbitrary, we conclude that
\begin{equation}
\label{taucons}
D_i \tau^{ij} =0;
 \end{equation}
i.e., $\tau^{ij}$ is covariantly conserved on $\partial M$ when the equations of motion hold in the bulk. In fact, since $\tau^{ij}$ is the radial conjugate momentum, it should be clear from chapter 17 that $\eqref{taucons}$ can also be derived directly from the equations of motion by evaluating the radial-version of the \ind{diffeomorphism constraint} on $\partial M$.  (The radial version of the Hamiltonian constraint imposes another condition on $\tau^{ij}$ that can be used to determine the trace $\tau = \tau^{ij} h_{ij}$ in terms of the traceless part of $\tau^{ij}$.)

If we now take $\hat \xi$ to be a boundary Killing field, we find $D_i (\tau^{ij}\hat \xi_j) =0$, so that the so-called \ind{Brown-York charge}

\begin{equation}
\label{eq:BYC}
Q_{BY}[\xi] := - \int_C \sqrt{q} \, n_i \tau^{ij} \hat \xi_j
\end{equation}
is independent of the choice of Cauchy surface $C$ in $\partial M$.  Here $n_i$ is a unit future-pointing normal to $C$ and $\sqrt{q}$ is the volume element induced on $C$ by $h_{ij}$.  Although these charges were defined by methods quite different from the Hamiltonian techniques of chapter 17, we will argue in section \ref{Hamlink} below that the end result is identical up to a possible choice of zero-point.  Once again, the argument will turn out to be essentially the same as one would give for field theory in a fixed non-dynamical background.

Before proceeding to the AdS case, let us take a moment to consider other possible boundary conditions.  We see from \eqref{eq:varyDIB} that the action \eqref{eq:DIB} also defines a valid variational principle for the boundary condition $\tau^{ij} =0$.  Of course, with this choice the charges \eqref{eq:BYC} all vanish.  But this should be no surprise.  Since the condition
$\tau^{ij} =0$ is invariant under {\it all} diffeomorphisms of $M$, there is no preferred subset of non-trivial asymptotic symmetries; all diffeomorphisms turn out to generate pure gauge transformations.  One may also study more complicated boundary conditions by adding additional boundary terms to the action \eqref{eq:DIB}, though we will not pursue the details here.

\subsection{Variational principles for scalar fields in AdS}
\label{sec:scalarS}

As the reader might guess, our discussion of AlAdS gravity will follow in direct analogy to the above treatment of gravity in a box.  Indeed, the only real difference is that we must work a bit harder to construct a good variational principle.  We will first illustrate the  relevant techniques below by constructing a variational principle for a \subind{scalar field on a fixed AdS background}{Variational Principle}, after which we will apply essentially identical techniques to AdS gravity itself in section \ref{sec:CTs}.

We will construct our variational principle using the so-called  counterterm subtraction approach pioneered in~\cite{Henningson:1998gx,Balasubramanian:1999re} and further developed in~\cite{deHaro:2000xn,Skenderis:2002wp}.  Our discussion below largely follows \cite{Skenderis:2002wp}, with minor additions from \cite{Andrade:2011dg}. We begin with the bulk action $S^\mathrm{Bulk}_\phi$ of
\eqref{eq:scalarS} and compute
\begin{equation}
\label{eq:dSp}
\delta S^\mathrm{Bulk}_\phi = -\int_{\partial M} \sqrt{|h|} \hat r^\mu \partial_\mu \phi \delta\phi,
\end{equation}
where $\hat r^\mu$ is the outward-pointing unit normal to $\partial M$ so that $\hat r^\mu \partial_\mu = - \frac{z}{\ell}\partial_z$.  The form of \eqref{eq:dSp} might appear to suggest that $S^\mathrm{Bulk}_\phi$ defines a good variational principle for any boundary condition that fixes $\phi$ on $\partial M$. But the appearance of inverse powers of $z$ means that we must be more careful, and that $S^\mathrm{Bulk}_\phi$ will suffice only when $\delta \phi$ vanishes sufficiently rapidly.

It is therefore useful to write \eqref{eq:dSp} in terms of the finite coefficients $\phi^{(2n)}, \phi^{(2(\nu+n))}$ of \eqref{eq:scalarexpansionGen} (or the corresponding coefficients in \eqref{eq:scalarexpansionInt}).  The exact expression is not particularly enlightening, and for large $\nu$ there are many singular terms to keep track of.  What is useful to note however is that all of the singular terms turn out to be exact variations.  In particular, using \eqref{eq:p2} one may show for non-integer $\nu < 2$ that the action

\begin{equation}
\label{eq:SpCT}
S_\phi = S^\mathrm{Bulk}_\phi + \int_{\partial M} \sqrt{|h|} \left( - \frac{ \Delta_- }{2 \ell}  \phi^2 + \frac{\ell}{4(\nu -1)} h^{ij} \partial_i \phi \partial_j \phi \right)
\end{equation}
satisfies
\begin{equation}
\label{eq:vSp}
\delta S_\phi = 2 \nu \ell^{d-1} \int_{\partial M} \sqrt{|\gamma^{(0)}|} \phi^{(2\nu)}\delta \phi^{(0)}.
\end{equation}
Since the boundary terms in \eqref{eq:SpCT} are each divergent in and of themselves, they are known as \ind{counterterms} in analogy with the counterterms used to cancel ultraviolet divergences in quantum field theory.  These divergences cancel against divergences in $S_{\phi}^\mathrm{Bulk}$ and the full action $S_\phi$ is finite for any field of the form \eqref{eq:scalarexpansionGen} with non-integer $\nu < 2$.  Similar results hold for non-integer $\nu > 2$ if additional higher-derivative boundary terms are included in \eqref{eq:SpCT}.  We will comment on differences for integer $\nu$ at the end of this section.

It is clear that $S_\phi$ provides a good variational principle so long as the boundary conditions either fix $\phi^{(0)}$ or set $\phi^{(2\nu)}=0$.  We may now identify
\begin{equation}
\label{eq:Phibndy}
\Phi_\mathrm{bndy} : = 2 \nu \ell^{d-1} \phi^{(2\nu)}
\end{equation}
as an AdS scalar \ind{response function} analogous to the boundary stress tensor $\tau^{ij}$ introduced in section \ref{sec:DIB}.  Note that adding an extra boundary term $\int \sqrt{\gamma^{(0)}}W[\phi^{(0)}]$ to $S_\phi$ allows one to instead use the Robin-like boundary condition
\begin{equation}
\label{eq:Wp}
\phi^{(2\nu)} = - \frac{\ell }{ 2 \nu} W'[\phi^{(0}],
\end{equation}
where $W'$ denotes the derivative of $W$ with respect to its argument.

Recall from section  \ref{sec:GravwithMatter} that requiring the energy to be bounded below restricts $\nu$ to be real (in which case we take $\nu$ non-negative).  That there are further implications for large $\nu$ can also be seen from \eqref{eq:SpCT}.  Note that the final term in \eqref{eq:SpCT} is a kinetic term on $\partial M$ and that for $\nu > 1$ it has a sign {\it opposite} to that of the bulk kinetic term.  Counting powers of $z$ shows that this boundary kinetic term vanishes at $\partial M$ for $\nu < 1$, but contributes for $\nu > 1$.  In this case, for any perturbation that excites $\phi^{(0)}$ and which is supported sufficiently close to $\partial M$, the boundary kinetic term in \eqref{eq:SpCT} turns out to be more important than the bulk kinetic term.  Thus the perturbation has negative kinetic energy.    One says that the theory contains \ind{ghosts}, and any conserved energy is expected to be unbounded below \cite{Andrade:2011dg}.  For this reason, for $\nu > 1$ one typically allows only \ind{boundary conditions} that fix $\phi^{(0)}$.  Of course, as noted in section \ref{sec:scalarS}, for $\nu > d/2$ coupling the theory to dynamical gravity and requiring the spacetime to be AlAdS will further require $\phi^{(0)} =0$.  On the other hand, for real $0< \nu < 1$ all of the above boundary conditions lead to ghost-free scalar theories.

The story of non-integer $\nu > 2$ is much the same as that of $\nu \in (1,2)$.  Adding additional higher-derivative boundary terms to \eqref{eq:SpCT} again leads to an action that satisfies \eqref{eq:vSp}.  While one can find actions compatible with general boundary conditions \eqref{eq:Wp}, the only ghost-free theories fix $\phi^{(0)}$ on $\partial M$. The story of integer $\nu$ is more subtle; the factors of $\ln z$ arising in that case from \eqref{eq:scalarexpansionInt} mean that we can find a good variational principle only by including boundary terms that depend explicitly on the defining function $\Omega$ of the chosen conformal frame.  Doing so again leads to ghosts unless $\phi^{(0)}$ is fixed as a boundary condition \cite{Andrade:2011dg}.

\subsection{A variational principle for AlAdS gravity}
\label{sec:CTs}

We are now ready to construct our variational principle for \subind{AlAdS gravity}{Variational Principle}. As for the scalar field above, we will start with a familiar bulk action and then add boundary terms.  One may note that in the scalar case our final action \eqref{eq:SpCT} consists essentially of adding boundary terms to  $S^\mathrm{Bulk}_\phi$ which i) are written as integrals of local scalars built from $\phi$ and its tangential derivatives along $\partial M$ and ii) precisely cancel divergent terms in $S^\mathrm{Bulk}_\phi$.  This motivates us to follow the strategy of~\cite{deHaro:2000xn} for the gravitational case in which we first identify divergent terms in a familiar action and write these terms as local scalars on $\partial M$.  We may then  construct a finite so-called renormalized action by adding boundary \ind{counterterms} on $\partial M$ to cancel the above divergences.  At the end of this process we may check that this renormalized action yields a good variational principle for interesting boundary conditions.  In analogy with section \eqref{sec:DIB}, for simplicity in the remainder of this chapter  we take the induced (conformal) metric on $\partial M$ to be globally hyperbolic with compact Cauchy surfaces.

Let us begin with an action containing the standard Einstein-Hilbert and cosmological constant terms in the bulk, along with the Gibbons-Hawking term.  It will facilitate our discussion of divergent terms to consider a regulated action in which the boundary has effectively been moved in to $z=\epsilon $.  For the moment, we choose some $\epsilon_0 > \epsilon$ and impose the Fefferman-Graham gauge \eqref{eq:FG} for all $z < \epsilon_0 $, so that this gauge holds in particular at the regulated boundary.    This gauge fixing at finite $z$ is merely an intermediate step to simplify the analysis.  We will be able to loosen this condition once we have constructed the final action. We let $h_{ij} = (\ell/z)^2 \gamma_{ij}|_{z=\epsilon  }$ be the induced metric on this regulated boundary and study the action
\begin{align} \label{Sreg}
S_\mathrm{reg} &= \frac{1}{2\kappa } \int_{z\ge \epsilon  } \sqrt{|g|} (R\edit{-}2 \Lambda) - \frac{1}{\kappa}\int_{z=\epsilon } \sqrt{|h|} K  \\
&= \edit{-}\frac{\ell^{d-1} }{2\kappa } \int_{z = \epsilon  } \sqrt{|{\gamma}^{(0)}| } \left( \epsilon^{-d} a_{(0)} +  \epsilon^{-d+2} a_{(2)} + \dots + \epsilon^{-2} a_{(d-2)} - \log(\epsilon^2) a_{(d)} \right) + (\mathrm{finite}), \nonumber
\end{align}
where $K= h_{ij} K^{ij}$ is the trace of the extrinsic curvature of the regulated boundary $\partial M_\epsilon$ at $z=\epsilon $ and the form of the divergences follows from \eqref{eq:FGexpansion}.  The coefficient $a_{(d)}$ vanishes for odd $d$.  For even $d$ it is called the \ind{conformal anomaly} for reasons to be explained below.

In analogy with the scalar field results of section \ref{sec:scalarS}, one finds that the coefficients $a_{(n)}$ which characterize the divergent terms are all local scalars built from $\gamma^{(0)}_{ij}$ and its derivatives along $\partial M$.  This follows directly from the fact that all terms $\gamma^{(n)}$ with $n \le d$ in the Fefferman-Graham expansion \eqref{eq:FGexpansion} are local functions of $\gamma^{(0)}_{ij}$ and its derivatives along $\partial M$.  Dimensional analysis shows that $a_{(n)}$ involves precisely $2n$ derivatives and the detailed coefficients $a_{(n)}$ can be found to any desired order by direct calculation.  For example, for $n \ne d$ the $a_{(n)}$ are given by (see e.g.~\cite{deHaro:2000xn})
\begin{align}
\label{eq:an}
& a_{(0)} =  - 2(d-1),  \ \ \ a_{(2)} = \frac{(d-4)}{2(d-2)} {\cal R}, \cr
& a_{(4)}= -\frac{d^2 - 9d + 16}{4(d-4)} \left( \frac{d {\cal R}^2 }{4(d-2)^2(d-1)} - \frac{ {\cal R}^{ij} {\cal R}_{ij} }{(d-2)^2} \right), \ \ \ \dots,
\end{align}
where as in section \ref{sec:FG}, ${\cal R}$ and ${\cal R}_{ij}$ are the Ricci scalar and Ricci tensor of $\gamma^{(0)}$ on $\partial M$.  For $d=2$,~$4$, the $\log$ terms are given by
\begin{align}
d = 2:& \ \ a_{(2)} = \edit{-}\frac{ {\cal R} }{2} , \cr
d = 4:& \ \ a_{(4)}= \left( \frac{  {\cal R}^2 }{24} - \frac{ {\cal R}^{ij} {\cal R}_{ij} }{8} \right).
\end{align}

As foreshadowed above, we now define the renormalized action
\begin{align}
S_\mathrm{ren} = \lim_{\epsilon\rightarrow 0} \left( S_\mathrm{reg} + S_\mathrm{ct} \right) ,
\end{align}
where
\begin{align} \label{eq:Sctgamma}
S_\mathrm{ct} := \frac{\ell^{d-1} }{2\kappa }\int_{z = \epsilon } \sqrt{-{\gamma}^{(0)}} \left( \epsilon^{-d} a_{(0)} +  \epsilon^{-d+2} a_{(2)} + \dots + \epsilon^{-2} a_{(d-2)} - \log(\epsilon^2) a_{(d)} \right)
\end{align}
is constructed to precisely cancel the divergent terms in $S_\mathrm{ren}$.  The representation \eqref{eq:Sctgamma} makes the degree of divergence in each term manifest.  But the use of $\epsilon$ in defining $S_\mathrm{ct}$ suggests a stronger dependence on the choice of defining function $\Omega$ (and thus, on the choice of conformal frame) than is actually the case.  To understand the true dependence, we should use the Fefferman-Graham expansion to instead express $S_\mathrm{ct}$ directly in terms of the (divergent) metric $h$ induced on $\partial M$ by the unrescaled bulk metric $g$ as was done in \cite{Balasubramanian:1999re}.  Dimensional analysis and the fact that each $a_{(n)}$ involves precisely $2n$ derivatives shows that this removes all explicit dependence on $\epsilon$ save for the logarithmic term in even $d$.  In particular, formally taking $\epsilon$ to zero we may write
\begin{align}
\label{eq:Sctg}
S_\mathrm{ct}  = \frac{ \ell }{2\kappa } \int_{\partial M}  \sqrt{ |h| } \left[ -\frac{2 (d-1)}{\ell^{2}} \edit{-} \frac{ {\cal R}_h }{(d-2)} + \dots - \frac{\epsilon^d \log(\epsilon^2) a_{(d)} }{\ell^2} \right],
\end{align}
where the ${\cal R}_h$ (Ricci scalar of $h$) term only appears for $d\ge 3$ and the dots represent additional terms that appear only for $d\ge 5$.

In general, the coefficients in \eqref{eq:Sctg} differ from those in \eqref{Sreg} due to sub-leading divergences in a given term in \eqref{eq:Sctg} contributing to the coefficients of seemingly lower-order terms in \eqref{Sreg}.  But the logarithmic term has precisely the same coefficient $a_{(d)}$ in both \eqref{eq:Sctg} and \eqref{Sreg}. Since the logarithmic term in \eqref{eq:FGexpansion} is multiplied by $z^d$, only the leading
 $ -\frac{2(d-1)}{\ell^{2}}\sqrt{|h|}$ term in \eqref{eq:Sctg} could contribute to any discrepancy.  But the first variation of a determinant is a trace, and the trace of the logarithmic coefficient $\bar \gamma^{(d)}_{ij}$ vanishes by \eqref{eq:gbd}.

Thus for $d$ odd (where the log term vanishes) the renormalized action $S_{\mathrm{ren}}$ can be expressed in a fully covariant form in terms of the physical metric $g$; all dependence on the defining function $\Omega$ (and so on the choice of conformal frame) has disappeared.  We therefore now drop the requirement that any Fefferman-Graham gauge be imposed for odd $d$.  But for even $d$, the appearance of $\log(\epsilon^2)$ in \eqref{eq:Sctg} indicates that $S_{\mathrm{ren}}$ does in fact depend on the choice of defining function $\Omega$ (and thus on the choice of conformal frame).  In analogy with quantum field theory, this dependence is known as the \ind{conformal anomaly}.  By replacing $\epsilon$ with $\Omega$ in \eqref{eq:Sctg}, we could again completely drop the requirement of Fefferman-Graham gauge in favor of making explicit the above dependence on $\Omega$.  However, an equivalent procedure is to require that the expansion \eqref{eq:FGexpansion} hold up through order $\gamma^{(d)}$ and to replace $\epsilon$ in \eqref{eq:Sctg} by the Fefferman-Graham coordinate $z$.  We will follow this  latter approach (which is equivalent to imposing Fefferman-Graham gauge only on the stated terms in the asymptotic expansion) as it is more common in the literature.

We are finally ready to explore variations of $S_\mathrm{ren}$. Since $S_\mathrm{ren}$ was constructed by adding only boundary terms to the usual bulk action, we know that $\delta S_\mathrm{ren}$ must be a pure boundary term on solutions.  As before, we will discard boundary terms in the far past and future of $M$ and retain only the boundary term at $\partial M$. Since $\partial M$ is globally hyperbolic with compact Cauchy surfaces, performing integrations by parts on $\partial M$ will yield boundary terms only in the far past and future of $\partial M$.  Discarding these as well allows us to write
\begin{equation}
\label{eq:vSren1}
\delta S_{\mathrm{ren}} = \int_{\partial M} S^{\mu \nu} \delta g_{\mu \nu},
\end{equation}
for some $S^{\mu \nu}$.  But let us now return to Fefferman-Graham gauge and use it to expand $\delta g_{\mu \nu}$ as in \eqref{eq:FGexpansion}.  Since $S_{\mathrm{ren}}$ is finite, $\delta S_{\mathrm{ren}}$ must be finite as well.  But the leading term in $\delta g_{\mu \nu}$ is of order $z^{-2}$.  So the leading term in $S_{\mu \nu}$ must be of order $z^2$.  It follows that only these leading terms can contribute to \eqref{eq:vSren1}.  Since the leading term in $\delta g_{\mu \nu}$ involves $\delta \gamma^{(0)}_{ij}$, we may write
\begin{equation}
\label{eq:vSren2}
\delta S_{\mathrm{ren}} = \frac{1}{2}\int_{\partial M} \sqrt{|\gamma^{0}|} \ T^{ij}_\mathrm{bndy} \delta \gamma^{(0)}_{ij}
\end{equation}
for some finite so-called \subind{boundary stress tensor}{stress tensor} $T^{ij}_\mathrm{bndy}$ on $\partial M$.  For odd $d$, the fact that $S_{\mathrm{ren}}$ is invariant under arbitrary changes of conformal frame $\delta \gamma^{(0)}_{ij} = e^{-2\sigma} \gamma^{(0)}_{ij}$  immediately implies that the boundary stress tensor is traceless: $T_\mathrm{bndy} := \gamma^{(0)}_{ij} T^{ij}_\mathrm{bndy} =0$.  In even dimensions, the trace is determined by the conformal anomaly of $S_{\mathrm{ren}}$ (i.e., by the logarithmic term in either \eqref{Sreg} or \eqref{eq:Sctg}) and one finds
\begin{equation}
\edit{T_\mathrm{bndy} = - \frac{\ell^{d-1}}{\kappa} \, a_{(d)}.}
\end{equation}
This result may also be derived by considering the radial version of the Hamiltonian constraint from chapter 17 and evaluating this constraint at $\partial M$.

Comparing with section \ref{sec:DIB}, it is clear that we may write
\begin{equation}
T_\mathrm{bndy}^{ij} = \lim_{\epsilon \rightarrow 0}  \left(\frac{\ell}{\epsilon}\right)^{d+2} \left( \tau^{ij} + \tau_{ct}^{ij}\right),
\end{equation}
where again $\tau_{ij} = \kappa^{-1} (K_{ij} - K h_{ij})$ and the new term $\tau^{ij}_\mathrm{ct}$ comes from varying $S_\mathrm{ct}$. In Fefferman-Graham gauge one finds by explicit calculation that
for $d$ odd
\begin{equation}
\label{eq:Tdodd}
T_\mathrm{bndy}^{ij} = \frac{d\ell^{d-1}}{2\kappa} {\gamma^{(d)}}^{ij}.
\end{equation}
For $d$ even there are extra contributions associated with the conformal anomaly, which are thus all determined by $\gamma^{(0)}$; e.g. (see~\cite{deHaro:2000xn})
\begin{align}
\label{eq:Texp}
{\rm for} \ d=2: \quad & T_\mathrm{bndy}^{ij} = \frac{\ell}{\kappa} \left({\gamma^{(2)}}^{ij} \edit{+\frac{1}{2}} {\cal R} {\gamma^{(0)}}^{ij} \right)  \\
{\rm for} \ d=4: \quad & T_\mathrm{bndy}^{ij} = \frac{2 \ell^3}{\kappa} \left[ { \gamma^{(4)} }^{ij}  - \frac{1}{8} \left( (\gamma^{(2)})^2 - { \gamma^{(2)} }^{kl} \gamma^{(2)}_{kl} \right) {\gamma^{(0)}}^{ij} \right. \cr
& \qquad \qquad  \qquad \left. - \frac{1}{2} { \gamma^{(2)} }^{ik} { { \gamma^{(2)} }_k }^j + \frac{1}{4} \gamma^{(2)} { \gamma^{(2)} }^{ij} + \frac{3}{2}  \bar{\gamma}^{(4)} {}^{ij} \right],
\end{align}
where $\gamma^{(2)}$, $\bar \gamma^{(4)}$ are given by \eqref{eq:g2}, \eqref{eq:gb2}, \eqref{eq:gb4}.
In all cases, we see that we may use $\gamma^{(0)}_{ij}, T_\mathrm{bndy}^{ij}$ to parametrize the free data in the Fefferman-Graham expansion.

The reader should note that the particular value of $T^{ij}_\mathrm{bndy}$ on a given solution depends on the choice of a representative $\gamma^{(0)}$ and thus on the choice of conformal frame.  For $d$ odd this dependence is a simple scaling, though it is more complicated for $d$ even.

But this does not diminish the utility of $T_\mathrm{bndy}^{ij}$.  For example, we see immediately from \eqref{eq:vSren2} that $S_{\mathrm{ren}}$ defines a good variational principle whenever i) $\gamma^{(0)}$ is fixed as a \ind{boundary condition} or ii) $d$ is odd, so that $T^{ij}_\mathrm{bndy}$ is traceless, and we fix only the conformal class of $\gamma^{(0)}$.

We close this section with some brief comments on other possible boundary conditions.  We see from \eqref{eq:vSren2} that $S_{\mathrm{ren}}$ is also a good variational principle if we fix $T_\mathrm{bndy}^{ij} =0$.  As in section \ref{sec:scalarS}, one may obtain variational principles for more complicated boundary conditions by adding further finite boundary terms to \eqref{eq:Sctg}; see \cite{Compere:2008us} for details.  However, just as for scalar fields with $\nu > 1$, boundary conditions that allow $\gamma^{(0)}$ to vary generally lead to ghosts \cite{Andrade:2011dg} (with the exception that, for $d$ odd no ghosts arise from allowing $\gamma^{(0)}$ to vary by a conformal factor).  For this reason we consider below only boundary conditions that fix $\gamma^{(0)}$, or at least its conformal class for $d$ odd.

\subsection{Conserved Charges for AlAdS gravity}
\label{sec:Qtime}

We are now ready to apply the Brown-York-type procedure discussed in section \ref{sec:DIB} to construct conserved charges for AlAdS gravity. The key step is again an argument analogous to \eqref{eq:BYcons} to show conservation of $T_\mathrm{bndy}^{ij}$ on $\partial M$.  We give the derivation here in full to highlight various subtleties of the AdS case.  We also generalize the result slightly by coupling the AlAdS gravity theory of section \ref{sec:CTs} to the scalar theory of section \ref{sec:scalarS}.   For definiteness we assume that the boundary conditions fix both $\gamma^{(0)}$ and $\phi^{(0)}$ (up to conformal transformations $(\gamma^{(0)}_{ij}, \phi^{(0)}) \rightarrow (e^{-2\sigma} \gamma^{(0)}_{ij}, e^{\Delta_- \sigma}\phi^{(0)})$)  for odd $d$, where the transformation of $\phi^{(0)}$ is dictated by \eqref{eq:scalarexpansionGen} and we take $\nu$ non-integer so that no log terms arise from the scalar field.  However, the more general case is quite similar \cite{Hollands:2005ya,Compere:2008us}.

We thus consider the action $S_\mathrm{total} = S_{\mathrm{ren}} + S_\phi$. The reader should be aware that, because the counterterms in $S_\phi$ explicitly depend on the boundary metric $\gamma^{(0)}$, this coupling to matter will change certain formulae in section \ref{sec:CTs}.  In particular, if we now make the natural definition
\begin{equation}
\label{eq:bstwp}
T_\mathrm{bndy}^{ij} =  \frac{2}{\sqrt{|\gamma^{(0)}|}} \frac{\delta S_\mathrm{total}}{\delta \gamma_{ij}^{(0)}},
\end{equation}
varying the action under a boundary conformal transformation leads to the more general condition
\begin{equation}
\label{eq:trwp}
T_\mathrm{bndy}  - \Delta_- \Phi_\mathrm{bndy} \phi^{(0)}=  - \frac{\ell^{d-1} a_{(d)}}{\kappa},
\end{equation}
which reduces to the trace constraint of section \ref{sec:CTs} only for $\Phi_\mathrm{bndy} =0$,  $\phi^{(0)} =0$, or $\Delta_- =0.$  Recall that $\Phi_\mathrm{bndy}$ is given by \eqref{eq:Phibndy}.

The coupling to $S_\phi$ similarly modifies the divergence condition \eqref{eq:BYcons} of section \ref{sec:DIB}.    Using the definition  \eqref{eq:bstwp}, we find
\begin{equation}
\label{eq:vSt}
\delta S_\mathrm{total} = \int_{\partial M} \sqrt{|\gamma^{(0)}|} \left( \frac{1}{2}T^{ij}_\mathrm{bndy} \delta \gamma^{(0)}_{ij} + \Phi_\mathrm{bndy} \delta \phi^{(0)}\right).
\end{equation}
Let us consider the particular variation associated with a bulk diffeomorphism $\xi$.  It is sufficient here to consider bulk diffeomorphisms compatible with whatever defining function $\Omega$ we have used to write \eqref{eq:vSt}; i.e., for which $\pounds_\xi \Omega =0$.  As described in section \ref{sec:diffeos},  other diffeomorphisms differ only in that they also induce a change of conformal frame.  Since we already extracted the information about $T_\mathrm{bndy}^{ij}$ (and in particular, about its trace) that can be obtained by changing conformal frame in section \ref{sec:CTs}, we lose nothing by restricting here to vector fields with $\pounds_\xi \Omega =0$.

As described in section \ref{sec:diffeos}, we then find $\delta \gamma^{(0)} = \pounds_{\hat \xi} \gamma^{(0)}$, $\delta \phi^{(0)} = \pounds_{\hat \xi} \phi^{(0)}$, where $\hat \xi$ is the vector field induced by $\xi$ on $\partial M$.     Thus \eqref{eq:vSt} reads
\begin{align} \label{eq:TijInvariance}
\delta_\xi S_\mathrm{ren} = 0 &=  \int_{ \partial M } \sqrt{|{\gamma}^{(0)}|} \left(T^{ij} D_i \hat \xi_j + \frac{\delta S_\mathrm{ren}}{\delta \phi^{(0)} } \pounds_{\hat \xi} \phi^{(0)} \right) \cr
&=  -\int_{ \partial M } \sqrt{|{\gamma}^{(0)}|}  \hat \xi_j \left(  D_i T^{ij} -  \Phi_\mathrm{bndy} D^j \phi^{(0)} \right),
\end{align}
where $D_i$ is again the covariant derivative on $\partial M$ compatible with with $\gamma^{(0)}$, all indices are raised and lowered with $\gamma^{(0)}$, and we have dropped the usual surface terms in the far past and future of $\partial M$.  Recalling that all $\hat \xi^i$ can arise from bulk vector fields $\xi$ compatible with any given $\Omega$, we see that \eqref{eq:TijInvariance} must hold for any $\hat \xi_j$.  Thus,
\begin{align} \label{eq:TijBalance}
D_i T^{ij}_\mathrm{bndy} = \Phi_\mathrm{bndy}   D^j \phi^{(0)};
\end{align}
 i.e., $T^{ij}_\mathrm{bndy}$ is conserved on $\partial M$ up to terms that may be interpreted as scalar sources.    These sources are analogous to sources for the stress tensor of, say, a scalar field on a fixed spacetime background when the scalar field is also coupled to some background potential.  Here the role of the background potential is played by $\phi^{(0)}$, which we have fixed as a boundary condition.  As in section \ref{sec:DIB}, the divergence condition \eqref{eq:TijBalance} may also be derived from the radial version of the \ind{diffeomorphism constraint} from chapter 17 evaluated on $\partial M$.  For $\phi^{(0)} =0$ and $d$ odd one immediately arrives at  \eqref{dodd}  using \eqref{eq:TijBalance} and \eqref{eq:Tdodd}.

We wish to use \eqref{eq:TijBalance} to derive conservation laws for asymptotic symmetries.  Here it is natural to say that a diffeomorphism $\xi$ of $M$ is an asymptotic symmetry if the there is {\it some} conformal frame in which the induced vector field $\hat \xi$ on $\partial M$ is i) a Killing field of $\gamma^{(0)}$  and ii) a solution of $\pounds_{\hat \xi} \phi^{(0)} = 0$. Due to the transformations of $\gamma^{(0)}, \phi^{(0)}$ under boundary conformal transformations, this is completely equivalent to first choosing an arbitrary conformal frame and then requiring
\begin{equation}
\label{eq:AsGCF}
\pounds_{\hat \xi} \gamma^{(0)}_{ij} = -2\sigma \gamma^{(0)}_{ij}, \ \ \
\pounds_{\hat \xi} \phi^{(0)} = \Delta_- \sigma \phi^{(0)}.
\end{equation}
The first requirement says that $\hat \xi$ is  a conformal Killing field of $\gamma^{(0)}_{ij}$ with $\frac{1}{d}D_i \hat \xi^i = - \sigma$ and the second says that it acts on $\phi^{(0)}$ like the corresponding infinitesimal conformal transformation.

For even $d$, we must also preserve the boundary condition that $\gamma^{(0)}$ be fixed (even including the conformal factor) and the requirement of section \eqref{sec:CTs} that Fefferman-Graham gauge hold to the first few orders in the asymptotic expansion.  An analysis similar to that of section \ref{sec:diffeos} then shows that we must have $\xi^z = \frac{z}{d}D_i \hat \xi^i$ to leading order near $\partial M$.  In particular, for $D_i \hat \xi^i \neq 0$ an asymptotic symmetry $\xi$ must be non-compatible with $\Omega$ is just the right way to leave $\gamma^{(0)}$ invariant.

As a side comment, we mention that the trivial asymptotic symmetries (the pure gauge transformations) are just those with $\hat \xi =0$.  This means that they act trivially on both $T_\mathrm{bndy}^{ij}$ and $\Phi_\mathrm{bndy}$ of section \ref{sec:scalarS}, so that both
both $T_\mathrm{bndy}^{ij}$ and the $\Phi_\mathrm{bndy}$ are \ind{gauge invariant}.  This conclusion is obvious in retrospect as these response functions are functional derivatives of the action with respect to the boundary conditions $\gamma^{(0)}_{ij}$ and $\phi^{(0)}$.  Since both the action and any boundary conditions are gauge invariant by definition, so too must be the functional derivatives $T_\mathrm{bndy}^{ij}$ and $\Phi_\mathrm{bndy}$.

Returning to our construction of \subind{charges}{AlAdS spacetimes}, note that for any asymptotic symmetries as above we may compute
\begin{equation}
\label{eq:loccons}
D_i (T_\mathrm{bndy}^{ij} \hat \xi_j) =  -\sigma( T_\mathrm{bndy}- \Delta_- \Phi_\mathrm{bndy}\phi^{(0)}   ) =   \sigma \frac{\ell^{d-1} a_{(d)}}{\kappa},
\end{equation}
where in the final step we have used \eqref{eq:trwp}.

In analogy with section \ref{sec:DIB}, we now consider the charges
\begin{align}
\label{eq:Q}
Q[\xi] &= -\int_{C} \sqrt{q} \, n_i T^{ij}_\mathrm{bndy} {\xi}_j,
\end{align}
where $C$ is a Cauchy surface of $\partial M$, $\sqrt{q}$ is the volume element induced on $C$ by $\gamma^{(0)}$, and $n^i$ is the unit future pointing normal to $C$ with respect to $\gamma^{(0)}$.  It follows from \eqref{eq:loccons} that these charges can depend on $C$ only through a term built from the conformal anomaly  $a_{(d)}$.

It is now straightforward to construct a modified charge $\tilde Q[\xi]$ which is completely independent of $C$.  The essential point here is to recall that $a_{(d)}$ depends only on the boundary metric $\gamma^{(0)}$.  Since we have fixed $\gamma^{(0)}$ as a boundary condition, the dependence on $C$ is the same for any two allowed solutions.  Thus on a given solution $s$ we need only define
\begin{equation}
\label{eq:tQ}
\tilde Q[\xi](s) = Q[\xi](s) -Q[\xi](s_0),
\end{equation}
where $s_0$ is an arbitrary reference solution satisfying the same boundary condition and which we use to set the zero-point.    The construction \eqref{eq:tQ} is sufficiently trivial that one often refers to $Q[\xi]$ itself as being conserved.

Our construction of the charges $Q[\xi]$, $\tilde Q[\xi]$ depended on the choice of some conformal frame. But it is easy to see that the charges are in fact independent of this choice for $d$ odd.  In that case, the factors $\sqrt{q}$, $n_i$, and $T_\mathrm{bndy}^{ij}$ all simply scale under a boundary conformal transformation and dimensional analysis shows that the combination \eqref{eq:Q} is invariant.   For even $d$ there are additional terms in the transformation of $T_\mathrm{bndy}^{ij}$.  But as usual these depend only on $\gamma^{(0)}$ so that they cancel between the two terms in \eqref{eq:tQ}.  Thus even in this case for fixed $s_0$ the charges \eqref{eq:tQ} are independent of the conformal frame.

To make the above procedure seem more concrete, we now quickly state results for the \ind{AdS${}_3$ and AdS${}_4$ Schwarzschild} solutions
\begin{align}
ds^2 = - \left(1 - \frac{2c_dGM}{\rho^{d-2}} + \frac{\rho^2}{\ell^2} \right) d\tau^2 + \frac{d\rho^2}{1 - \frac{2c_dGM}{\rho^{d-2}} + \frac{\rho^2}{\ell^2} } + \rho^2 d\Omega_{(d-2)}^2,
\end{align}
where $c_3=1$ and $c_4 = \frac{4}{3\pi}$.  The boundary stress tensor may be calculated by converting to Fefferman-Graham coordinates, say for the conformal frame defined by $\Omega = \rho^{-1}$.  (Note that the Fefferman-Graham radial coordinate $z$ will agree with $\rho$ only at leading order.)  One then finds the energy
\begin{align}
\label{AdSSM}
Q[-\partial_\tau] = \begin{cases}
M, & d=3 \\
M + \dfrac{3\pi \ell^2}{32 G}, & d=4,
\end{cases}
\end{align}
where we remind the reader that energies $E = -Q[\partial_\tau] = Q[-\partial_\tau]$ are conventionally defined in this way with an extra minus sign to make them positive. We see that for $d=3$ we recover the expected result for the energy of the spacetime.  For $d=4$ we also recover the expected energy up to a perhaps unfamiliar choice of zero-point which we will discuss further in section \ref{sec:algc}.

\subsection{Positivity of the energy in AlAdS gravity}
\label{sec:pos}

Thus far we have treated all charges $Q[\xi]$ on an equal footing.  But when $\hat \xi$ is everywhere timelike and future-directed on $\partial M$, it is natural to call $E=Q[-\xi]$ an \subind{{\it energy}}{asymptotically locally AdS} and to wonder if $E$ is bounded below.  Such a result was established in chapter 20 for the ADM energy of asymptotically flat spacetimes, and the Witten spinor methods  \cite{Witten:1981mf,Nester:1982tr} discussed there generalize readily to asymptotically AdS (AAdS) spacetimes so long as the matter fields satisfy the dominant energy condition and decay sufficiently quickly at $\partial M$ \cite{Townsend:1984iu}.  In particular, this decay condition is satisfied for the scalar field of section \ref{sec:scalarS} with $m^2 \ge m_{BF}^2$ when $\phi^{(0)}$ is fixed as a boundary condition.    Extensions to more general scalar boundary conditions can be found in \cite{Hertog:2005hm,Amsel:2006uf,Amsel:2007im,Faulkner:2010fh,Amsel:2011km}.  Here the details of the boundary conditions are important, as boundary conditions for which the $W$ of \eqref{eq:Wp} diverges sufficiently strongly in the negative direction tend to make any energy unbounded below (see e.g. \cite{Hertog:2006wj} for examples).  This is to be expected from the fact that, as discussed in section \ref{sec:scalarS}, this $W$ represents an addition to the Lagrangian and thus to any Hamiltonian, even if only as a boundary term.  As for $\Lambda =0$, the above AAdS arguments were inspired by earlier arguments based on quantum supergravity (see \cite{Deser:1977hu,Grisaru:1977gj} for the asymptotically flat case and \cite{Abbott:1981ff} for the AAdS case).

The above paragraph discussed only AAdS spacetimes.  While the techniques described there can also be generalized to many AlAdS settings, it is not possible to proceed in this way for truly general choices of $M$ and $\partial M$.  The issue is that the methods of \cite{Witten:1981mf,Nester:1982tr} require one to find a spinor field satisfying a Dirac-type equation subject to certain boundary conditions.  But for some $M, \partial M$ one can show that no solution exists.  In particular, this obstruction arises when $\partial M = S^1 \times {\mathbb R}^{d-1}$ and the $S^1$ is contractible in $M$ \cite{Horowitz:1998ha}.

The same obstruction also arises with zero cosmological constant in the context of Kaluza-Klein theories (where the boundary conditions may again involve an $S^1$ that is contractible in the bulk).  In that case, the existence of so-called bubbles of nothing demonstrates that the energy is in fact unbounded below and that the system is unstable even in vacuum \cite{Witten:1981gj,Brill:1989di}.  But what is interesting about the AlAdS context with $\partial M = S^1 \times {\mathbb R}^{d-1}$ is that there are good reasons \cite{Horowitz:1998ha} to believe that the energy {\it is} in fact bounded below -- even if there are there are some solutions with energy lower than what one might call empty AdS with $\partial M = S^1 \times {\mathbb R}^{d-1}$ (by which we mean the quotient of the Poincar\'e patch under some translation of the $x^i$).  Perhaps the strongest such argument (which we will not explain here) comes from AdS/CFT.  But another is that \cite{Witten:1998zw} identified a candidate lowest-energy solution (called the \ind{AdS soliton}) which was shown \cite{Horowitz:1998ha} to at least locally minimize the energy.  Proving that the AdS soliton is the true minimum of the energy, or falsifying the conjecture, remains an interesting open problem whose solution appears to require new techniques.

\section{Relation to Hamiltonian Charges}
\label{Hamlink}

We have shown that the charges \eqref{eq:tQ} are conserved and motivated their definition in analogy with familiar constructions for field theory in a fixed curved spacetime.  But it is natural to ask whether the charges \eqref{eq:tQ} in fact agree with more familiar Hamiltonian definitions of asymptotic charges constructed, say, using the AdS generalization of the Hamiltonian approach described in chapter 17. Denoting these latter charges $H[\xi]$, the short answer is that they agree so long as we choose $s_0$ in \eqref{eq:tQ} to satisfy $H[\xi](s_0)=0$; i.e., they agree so long as we choose the same (in principle arbitrary) zero-point for each notion of charge.  We may equivalently say that the difference $Q[\xi] - H[\xi]$ is the same for all solutions in our phase space, though for conformal charges it may depend on the choice of Cauchy surface $C$ for $\partial M$.  As above, for simplicity we take $\partial M$ to be globally hyperbolic with compact Cauchy surfaces.

This result may be found by direct computation (see \cite{Hollands:2005wt} for simple cases).  But a more elegant, more general,  and more enlightening argument can be given \cite{Hollands:2005ya} using a covariant version of the Poisson bracket known as the Peierls bracket \cite{Peierls}.  The essence of the argument is to show that $Q[\xi]$ generates the canonical transformations associated with the diffeomorphisms $\xi$.  This specifies all Poisson brackets of $Q[\xi]$ to be those of $H[\xi]$.  Thus $Q[\xi] - H[\xi]$ must be a c-number in the sense that all Poisson brackets vanish.  But this means that it is constant over the phase space.

After pausing to introduce the Peierls bracket, we sketch this argument below following \cite{Hollands:2005ya}.
As in section \ref{sec:Qtime}, we suppose for simplicity that the only bulk fields are the metric and a single scalar field with non-integer $\nu$ and we impose boundary conditions that fix both $\gamma^{(0)}_{ij}$ and $\phi^{(0)}$.  However, the argument for general bulk fields is quite similar \cite{Hollands:2005ya}.  While this material represents a certain aside from our main discussion, it will provide insight into the algebraic properties of conserved charges, the stress tensor itself, and a more general notion of so-called boundary observables that we will shortly discuss.

\subsection{The \ind{Peierls bracket}}
\label{pb}

The Peierls bracket is a Lie bracket operation that acts on
gauge-invariant functions on the space of solutions ${\cal S}$ of some theory.
As shown in the original
work \cite{Peierls}, this operation is equivalent to the Poisson
bracket under the natural identification of the phase space with
the space of solutions.  However, the
Peierls bracket is {\it manifestly} spacetime covariant.  In particular, one may directly define the Peierls bracket between any two quantities $A$ and $B$ located anywhere in spacetime, whether or not they may be thought of as lying on the same Cauchy surface.  In fact, both $A$ and $B$ can be highly non-local, extending over large regions of space and time. These features make the Peierls bracket ideal for studying the
boundary stress-tensor, which is well-defined on the space of
solutions but is not a local function in the bulk
spacetime.

To begin, consider two functions $A$ and $B$ on ${\cal S}$, which are in fact defined as functions on a larger space ${\cal H}$, which we call the space of histories.  This space ${\cal H}$ is the one on which the action is defined; i.e., the solution space
${\cal S}$ consists of those histories in ${\cal H}$ on which the action $S$ is stationary.  One may show that the Peierls bracket on ${\cal S}$ depends only on $A,B$ on ${\cal S}$ and not on their extensions to ${\cal H}$.

The Peierls bracket is defined by considering the effect on one gauge
invariant function (say, $B$)
when the action is deformed by a term proportional to another
such function ($A$).   One defines the advanced ($D^+_AB$)
and retarded ($D^-_AB$) effects of $A$ on $B$ by comparing the
original system with a new system given by the action
$S_{\epsilon} = S + \epsilon A$, but associated with the same
space of histories ${\cal H}$. Here $\epsilon$ is a real parameter which will
soon be taken to be infinitesimal, and the new action is
associated with a new space ${\cal S}_\epsilon$ of deformed
solutions.

\begin{figure}
\centering
\includegraphics[width=0.5\textwidth]{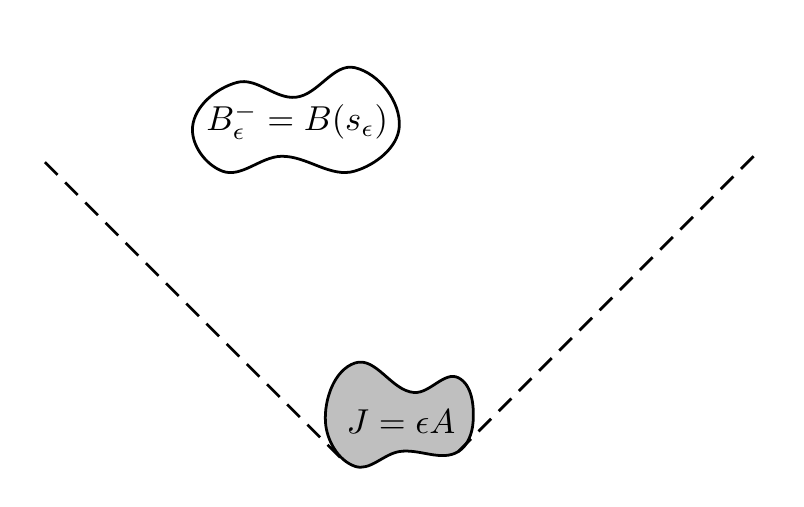}
\caption{An illustration of the definition of $B^-_\epsilon$.  A source term $J=\epsilon A$ is added to the action and the gauge invariant function $B$ is calculated for the deformed solution $s_\epsilon$ subject to the boundary conditions that $s$ and $s_\epsilon$ coincide in the far past.  Dashed lines indicate the boundary of the causal future of $J$.  Only functions $B$ which have support in this region can have $B(s_\epsilon)\ne B(s)$.  For visual clarity we have chosen our gauge invariant function $A$ and $B$ to have compact support though this is not required.}
\label{fig:Peierls}
\end{figure}

Under retarded (advanced) boundary conditions for which the
solutions $s \in {\cal S}$ and $s_{\epsilon} \in {\cal
S}_{\epsilon}$ coincide in the past (future) of the support
of $A$, the quantity $B_0 = B(s)$ computed using the undeformed
solution $s$ will in general differ from $B_{\epsilon}^\pm =
B(s_{\epsilon})$ computed using $s_{\epsilon}$ and retarded $(-)$
or advanced $(+)$ boundary conditions (see Fig.~\ref{fig:Peierls}).  For small epsilon, the
difference between these quantities defines the retarded
(advanced) effect $D^-_AB$ ($D^+_AB$) of $A$ on $B$ through:
\begin{equation}
\label{effects} D^{\pm}_AB = \lim_{\epsilon \rightarrow 0}
\frac{1}{\epsilon}(B_{\epsilon}^\pm-B_0^{} ),
\end{equation}
which is a function of the unperturbed solution $s$.  Similarly,
one defines $D^\pm_BA$ by reversing the roles of $A$ and $B$
above. Since $A,B$ are gauge invariant, $D^\pm_BA$ is a
well-defined (and again gauge-invariant) function on the space
${\cal S}$ of solutions so long as both $A$ and $B$ are
first-differentiable on ${\cal H}$.  This requirement may be
subtle if the spacetime supports of $A$ and $B$ extend into the far past and future, but is straightforward for objects like $T_\mathrm{bndy}^{ij}(x)$, $\Phi_\mathrm{bndy}(x)$ that are well-localized in time.

The Peierls
bracket \cite{Peierls} is then defined to be the difference of
the advanced and retarded effects:
\begin{equation}
\label{Peierls} \{ A,B \} = D^+_AB - D^-_AB.
\end{equation}

As shown in \cite{Peierls}, this operation agrees with the Poisson bracket (suitably generalized to allow $A,B$ at unequal times).  This  generalizes the familiar result that the commutator function for a free
scalar field is given by the difference between the advanced and
retarded Green's functions.  In fact, it is enlightening to write
the Peierls bracket more generally in terms of such Green's
functions.  To do so, let us briefly introduce the notation $\phi^I$ for a complete set of bulk fields (including the components of the bulk metric) and the associated advanced and retarded Green's
functions $G_{IJ}^\pm(x,x')$.  Note that we have
\begin{equation}
\label{GFs} D^+_AB = \int dx \ dx' \frac{\delta B}{\delta
\Phi^I(x)} G^+_{IJ}(x,x') \frac{\delta A}{\delta \Phi^J(x')} =
 \int dx \ dx' \frac{\delta B}{\delta \phi^j(x')} G^-_{JI}(x',x)
\frac{\delta A}{\delta \phi^j(x)} = D^-_BA,
\end{equation}
where we have used the identity $G^+_{IJ}(x,x') = G^-_{JI}(x',x)$.
Thus, the Peierls bracket may also be written in the manifestly
antisymmetric form
\begin{equation}
\label{Peierls2} \{ A,B \} = D^-_BA-D^-_AB  = D^+_AB-D^+_BA.
\end{equation}
The expressions (\ref{GFs}) in terms of $G^\pm_{IJ}(x,x')$ are
also useful in order to verify that the Peierls bracket defines a
Lie-Poisson algebra.  In particular, the derivation property
$\{A,BC\}= \{ A,B \} C + \{A,C\} B$ follows immediately from the
Leibnitz rule for functional derivatives.  The Jacobi identity
also follows by a straightforward calculation, making use of the
fact that functional derivatives of the action commute (see e.g.,
\cite{Bryce1,Bryce2}). If one desires, one may use related Green's
function techniques to extend the Peierls bracket to a Lie algebra
of gauge dependent quantities \cite{Marolf:1993af}.

\subsection{Main Argument}

\label{main}

We wish to show that the charges $Q[\xi]$ generate the appropriate asymptotic symmetry for any asymptotic Killing field $\xi$.  Since this is true by definition for any Hamiltonian charge $H[\xi]$, it will then follow that $Q[\xi]-H[\xi]$ is constant over the space of solutions ${\cal S}$.  We first address
the case where $\xi$ is compatible with $\Omega$, and then proceed to the more general case where
$\hat \xi$ acts only as a conformal Killing field on the
boundary.

Showing that $Q[\xi]$ generates diffeomorphisms along $\xi$ amounts to proving a certain version of \ind{Noether's theorem}. Recall that the proof of Noether's theorem involves examining the change in the action under a spacetime-dependent generalization of the desired symmetry.  The structure of our argument below is similar, where we consider both the action of a given asymptotic symmetry $\xi$ and the spacetime-dependent generalization $f \xi$ defined by choosing an appropriate scalar function $f$ on $M$.  It turns out to be useful to choose $f$ on $M$ (with restriction $\hat f$ to $\partial M$) such that
\begin{itemize}
\item  $f=0$ in the far past and $f=1$ in the far future.
\item $\hat f =0$ to the past of some Cauchy surface $C_0$ of $\partial
M$, and $\hat f=1$ to the future of some Cauchy surface $C_1$
of $\partial M$.
\end{itemize}

Suppose now that $\xi$ is an asymptotic symmetry compatible with
$\Omega$.  Then the bulk and boundary fields
transform as
\begin{equation}
\label{SymAct} \delta \phi = \pounds_{\xi}\phi, \quad \delta
g_{\mu \nu} = \pounds_\xi g_{\mu \nu}, \quad \delta \gamma^{(0)}_{ij} =
 \pounds_{\hat \xi} \gamma^{(0)}_{ij} = 0, \quad
{\rm and} \quad \delta \phi^{(0)} = \pounds_{\hat \xi}
\phi^{(0)}= 0.
\end{equation}
The key step of the argument is to construct a new transformation
$\Delta_{f,\xi}$ on the space of fields such that the associated
first order change $\Delta_{f,\xi} S$ in the action generates the
asymptotic symmetry $-\xi$.  We will first show that the above property turns out to hold for
\begin{equation}
\label{eq:D}
\Delta_{f,\xi}  := (\pounds_{f
\xi} - f\pounds_{\xi} ),
\end{equation}
and then verify that $\Delta_{f,\xi} S = -Q[\xi]$.
The form of $\Delta_{f,\xi} S$ is essentially that suggested in \cite{Sorkin:1986ph} using Hamilton-Jacobi methods, so our argument will also connect $Q[\xi]$ with \cite{Sorkin:1986ph}.

An important property of \eqref{eq:D} is that the changes $\Delta_{f,\xi} g_{\mu \nu}$ and $\Delta_{f,\xi} \phi$ are {\it
algebraic} in $\phi$ and $g_{\mu \nu}$; i.e., we need not take spacetime
derivatives of $g_{\mu \nu}, \phi$ to compute
the action of $\Delta_{f,\xi}$. Furthermore, $\Delta_{f,\xi} \phi$ and $\Delta_{f,\xi} g_{\mu \nu}$ are both
proportional to $\nabla_a f$, and so vanish in both the far future and the far past.  This guarantees that
$\Delta_{f,\xi} S$ is a differentiable function on ${\cal H}$.  In particular, solutions to the equations of motion resulting from the deformed action  $S + \epsilon \Delta_{f,\xi} S$ are indeed stationary points of  $S + \epsilon \Delta_{f,\xi} S$ under all
variations which preserve the conditions and vanish in the far future and past.

It is important to note that
the quantity $\Delta_{f,\xi} S$ is gauge-invariant when the equations of motion hold.  This is easy to see since by definition on ${\cal S}$ all variations of $S$ become pure boundary terms.  Boundary terms in the far past and future vanish due to the observations above, and since $\gamma^{(0)}_{ij}, \phi^{(0)}$ are fixed by boundary conditions the boundary terms on $\partial M$ depend on the bulk fields only through the gauge invariant quantities $T_\mathrm{bndy}^{ij}$ and $\Phi_\mathrm{bndy}$.   Thus, we may take the Peierls bracket of $\Delta_{f,\xi} S$ with
any other observable $A$.

We proceed by considering the modified action
\begin{equation}
\tilde S [\phi, g_{\mu \nu}] = S[\phi, g_{\mu \nu}] + \epsilon \Delta_{f,\xi} S[\phi, g_{\mu \nu}] = S[\phi + \epsilon \Delta_{f,\xi} \phi, g_{\mu \nu} + \epsilon \Delta_{f,\xi} g_{\mu \nu}],
\end{equation}
where the last equality holds to first order in $\epsilon$ (and in fact defines $\Delta_{f,\xi} S[\phi, g_{\mu \nu}]$).  Since $\tilde S$ is just $S$ with its argument shifted by $\epsilon \Delta_{f,\xi}$, the stationary points $s_1$ of $\tilde S$ are precisely the oppositely-shifted versions of the stationary points $s$ of $S$; i.e., we may write $s_1 = (1-\epsilon\Delta_{f,\xi}) s$ for some $s \in {\cal S}$.

We should of course ask if $s_1$ satisfies the desired boundary conditions on $\partial M$.
Since $\xi$ is compatible with $\Omega$, the
boundary fields shift in the same way as their bulk counterparts; i.e., those of $s_1$ have been shifted by $- \epsilon
\Delta_{f,\xi}$ relative to those of $s$.  Since $\xi$ is an asymptotic symmetry, its action preserves the boundary fields.  Now, the reader will note that there is a non-trivial effect from the $\pounds_{f\xi}$ term in $\Delta_{f,\xi}$.  But this term is a pure diffeomorphism, and since all boundary terms are covariant on $\partial M$ the action $\tilde S$ is invariant under {\it all} diffeomorphisms compatible with $\Omega$ (i.e., which preserve the given conformal frame), even those that act non-trivially on the boundary.  So the history
\begin{equation}
\label{s2} s_2 = (1 + \epsilon \pounds_{f \xi}) s_1
= (1 + \epsilon f \pounds_{\xi} ) s
\end{equation}
has
\begin{equation}
\label{s2bf} \phi^{(0)}|_{s_2} = \phi^{(0)}|_s, \quad
g_{\mu \nu}|_{s_2} = g_{\mu \nu}|_s,
\end{equation}
and again solves the equations of motion that follow from
$\tilde S$.

This observation allows a straightforward computation of the advanced and retarded
changes $D^\pm_{\Delta_{f,\xi}S}A$ for any gauge invariant quantity
$A$.  We first consider the retarded change evaluated on a solution $s$ as above.  We require a
solution $s^-_\epsilon$ of the perturbed equations of motion which
agrees with $s$ in the far past.  Since the infinitesimal
transformation $f\pounds_\xi$ vanishes in the far past, we may set $s_\epsilon^- = s_2$ as defined (\ref{s2}) above;
i.e. $s_\epsilon^- = (1+ \epsilon f\pounds_\xi) s$. Thus, the
retarded effect on $A$ is just $D^-_{\Delta_{f,\xi}S}A = f
\pounds_\xi A$.

To compute the advanced effect, we must find a solution $s^+_\epsilon$
of the perturbed equations of motion which agrees with $s$ in the far future.
Consider the history $s^+_\epsilon = (1 -
\epsilon\pounds_\xi)s^-_\epsilon = (1 + (f-1)\epsilon
\pounds_\xi)s$. Since this differs from $s^-_\epsilon$ by the
action of a symmetry compatible with $\Omega$, it again solves the desired
equations of motion (to first order in $\epsilon$) and induces the
required boundary fields (\ref{s2bf}).  In addition, $s^+_\epsilon$ and $s$ agree in the far future (where  $f=1$).
Thus, we may use $s^+_\epsilon$ to compute the advanced change in
any gauge invariant $A$:
\begin{equation}
D^+_{\Delta_{f,\xi}S}A = (f-1) \pounds_\xi A.
\end{equation}
Finally, we arrive at the Peierls bracket
\begin{equation}
\label{PBfinal} \{\Delta_{f,\xi}S   ,  A  \} =
D^+_{\Delta_{f,\xi}S } A - D^-_{\Delta_{f,\xi}S } A = -
\pounds_\xi A.
\end{equation}
As desired $-\Delta_{f,\xi}S$ generates a diffeomorphism along the
asymptotic symmetry $\xi$ as desired.

All that remains is to relate $\Delta_{f,\xi} S$ to $Q[\xi]$. But this
is straightforward.  Since $f$ vanishes in the far past and future we have
\begin{equation}
\label{DSvar} \Delta_{f,\xi} S =
 \int_M  \left(\frac{\delta S}{\delta \phi} \Delta_{f,\xi} \phi + \frac{\delta S}{\delta g_{\mu \nu}} \Delta_{f,\xi} g_{\mu \nu}\right) + \frac{1}{2} \int_{\partial M} \sqrt{\gamma^{(0)}} \, T^{ij}_\mathrm{bndy}\Delta_{f,\xi} \gamma_{ij}^{(0)} +
\int_{\partial M} \sqrt{\gamma^{(0)}} \, \Phi_\mathrm{bndy} \Delta_{f,\xi} \phi^{(0)}.
\end{equation}
But the bulk term vanishes on solutions $s \in {\cal S}$, and from \eqref{SymAct} we find $\Delta_{f,\xi} \phi^{(0)} =
(\pounds_{\hat f \hat \xi} -
\hat f\pounds_{\hat \xi} ) \phi^{(0)}  = 0$.
So only the term containing $T^{ij}_\mathrm{bndy}$ contributes to (\ref{DSvar}).

To compute the remaining term note that
\begin{equation}
\label{bound2}
\Delta_{f,\xi} \gamma^{(0)}_{ij} = (\pounds_{\hat f \hat \xi} -
\hat f\pounds_{\hat \xi} )  \gamma^{(0)}_{ij} = \hat{\xi}_i \partial_j \hat{f} + \hat{\xi}_j \partial_i \hat{f}.
\end{equation}
Since \eqref{bound2} vanishes when $f$ is constant, we may restrict the integral over $\partial M$ to the region $V$ between $C_0$ and $C_1$ and use the symmetry  $T^{ij}_\mathrm{bndy} = T^{ji}_\mathrm{bndy}$ to obtain
\begin{eqnarray}
\label{compatcalc}
\Delta_{f,\xi} S =
  &=&     \int_{{V}}  \sqrt{|\gamma^{(0)}|} T^{ij}_\mathrm{bndy}\xi_i \partial_j f \cr
  &=&  \int_{C_1}  \sqrt{q} \, n_j T^{ij}_\mathrm{bndy} \xi_i \,
 -  \int_{V} \sqrt{|\gamma^{(0)}|}  f D_i \left( T_\mathrm{bndy}^{ij}
\xi_j \right)
  \cr &=&  - Q_{C_1}[\xi].
\end{eqnarray}
Here we used the fact that $\hat f=0$ on $C_0$ to drop contributions from $C_0$ and the fact that that $\hat \xi$ is a Killing field of the boundary metric along with~\eqref{eq:loccons} to show that the $\int_V$ term in the second line vanishes.

Thus,  $-\Delta_{f,\xi} S$ agrees (on solutions) with the charge
$Q[\xi]$ evaluated on the cut $C_1$.  Since $Q[\xi]$ is conserved,  this equality also holds on any other cut of $\partial M$.  Having already shown by eq.~(\ref{PBfinal}) that the variation
$\Delta_{f,\xi} S$ generates the action of the infinitesimal
symmetry $-\xi$ on observables, it follows that $Q[\xi]$ generates the action of $\xi$:
\begin{equation}
\label{result} \{ Q[\xi], A \} = \pounds_\xi A,
\end{equation}
as desired.

\subsection{Asymptotic Symmetries not compatible with $\Omega$}

We now generalize the argument to asymptotic symmetries $\xi$ that are {\it
not} compatible with $\Omega$, so that $\hat \xi$ satisfies \eqref{eq:AsGCF}. The field content and boundary conditions are the same as above.  But the non-trivial action of $\xi$ on $\Omega$ means that there are now are additional terms when a diffeomorphism acts on the boundary fields $\phi^{(0)}, \gamma_{ij}^{(0)}$:

\begin{equation}
\label{bfs} \delta_{ \pounds_{f\xi}} \phi^{(0)} =
\pounds_{\hat f \hat \xi} \phi^{(0)}  - \Delta_- \hat f
\sigma \phi^{(0)} , \quad \delta_{ \pounds_{f\xi}} \gamma_{ij}^{(0)} =
\pounds_{\hat f \hat \xi} \gamma_{ij}^{(0)}  + 2\hat f \sigma \gamma_{ij}^{(0)}.
\end{equation}
Combining \eqref{eq:AsGCF} and \eqref{bfs} we see that $\delta_{\pounds_\xi}$ acts trivially on the boundary data $\gamma_{ij}^{(0)}, \phi^{(0)}$, as it must since asymptotic symmetries were defined to leave the boundary conditions invariant.  Thus the  histories
$s^\pm_\epsilon$ identified above (see, e.g.,
(\ref{s2}))  again satisfy the same boundary conditions as $s$.

In contrast to section \ref{main} the operation $\pounds_{f\xi}$ now acts non-trivially on $\Omega$ and thus on $S$.  But since this is only through the conformal anomaly term $a_{(d)}$ in \eqref{eq:Sctg}, $\pounds_{f\xi} S$ depends only on the boundary metric $\gamma^{(0)}$ and is otherwise constant on ${\cal H}$.  So the equations of motion are unchanged and the histories $s^\pm_\epsilon$ again solve the equations of motion for $\tilde S$.

It remains to repeat the analogue of the calculation \eqref{compatcalc}.  But here the only change is that the $\int_V$ term on the second line no longer vanishes. Instead, it  contributes a term proportional to $a_{(d)}$.  Since this term is constant on the space of solutions ${\cal S}$, it has vanishing Peierls brackets and we again conclude that
$Q_{C_1}[\xi]$ generates the asymptotic symmetry $\xi$. (This comment corrects a minor error in \cite{Marolf:1993af}.) And since $Q_{C}[\xi]$ depends on the Cauchy surface $C$ only through a term that is constant on ${\cal S}$, the same result holds for any $C$. Thus, even when $\hat \xi$ is only a conformal symmetry of the
boundary, $Q_C[\xi] - H[\xi]$ is constant over the space ${\cal S}$ of solutions.

\subsection{Charge algebras and central charges}
\label{sec:algc}

We saw above that our charges $Q[\xi]$ generate the desired asymptotic symmetries via the Peierls bracket.  This immediately implies what is often called the \ind{{\it representation theorem}}, that the algebra of the charges themselves matches that of the associated symmetries up to possible so-called central extensions.  This point is really quite simple.  Consider three vector field $\xi_1, \xi_2, \xi_3$ related via the Lie bracket through $\{\xi_1,\xi_2\} = \xi_3.$  Now examine the Jacobi identity

\begin{equation}
\label{eq:Jacobi}
\{ Q[\xi_1], \{Q[\xi_2], A\}\} + \{Q[\xi_2], \{A, Q[\xi_1]\}\} + \{A, \{Q[\xi_1], Q[\xi_2]\}\} = 0
\end{equation}
which must hold for any $A$.  Since $\{Q[\xi_i], B\} = \pounds_{\xi_i} B$ for any $B$, we may use \eqref{eq:Jacobi} to write
\begin{equation}
\pounds_{\xi_3}A =
\pounds_{\xi_1} \left( \pounds_{\xi_2} A \right) - \pounds_{\xi_2} \left( \pounds_{\xi_1} A \right) = \{\{Q[\xi_2], Q[\xi_1]\}, A\}.
\end{equation}
But the left-hand-side is also $\{Q[\xi_3], A\}$.  So we conclude that $\{Q[\xi_1], Q[\xi_2]\}$ generates the same transformation as $Q[\xi_3]$.  This means that they can differ only by some  $K(\xi_1,\xi_2)$ which is constant across the space of solutions (i.e., it is a so-called c-number):
\begin{equation}
\label{cext}
\{Q[\xi_1], Q[\xi_2]\} = Q[\{\xi_1,\xi_2\}] + K(\xi_1,\xi_2).
\end{equation}

For some symmetry algebras one can show that any such $K(\xi_i,\xi_j)$ can be removed by shifting the zero-points of the charges $Q[\xi]$.  In such cases the $K(\xi_i,\xi_j)$ are said to be trivial.  Non-trivial $K(\xi_i,\xi_j)$ are classified by a cohomology problem and are said to represent \ind{central extensions} of the symmetry algebra.

It is easy to show that $K(\xi_i,\xi_j)$ may be set to zero in this way whenever there is some solution (call it $s_0$) which is invariant under all symmetries.    The fact that it is invariant means that $\{Q[\xi_i], A\}(s_0) = 0$; i.e., the bracket vanishes when evaluated on the particular solution $s_0$ for any $\xi_i$ and any $A$.  So take $A = Q[\xi_j]$, and set the zero-points of the charges so that $Q[\xi](s_0) =0$.  Evaluating \eqref{cext} on $s_0$ then gives $K(\xi_i,\xi_j)(s_0) =0$ for all $\xi$.  But since $K(\xi_i,\xi_j)(s_0)$ is constant over the space of solutions this means that it vanishes identically.

For asymptotically flat spacetimes the asymptotic symmetries generate the Poincar\'e group, which are just the exact symmetries of Minkowski space.  Thus one might expect the asymptotic symmetries of $(d+1)$-dimensional AlAdS spacetimes to be (perhaps a subgroup of) $SO(d,2)$ in agreement with the isometries of AdS${}_{d+1}$ compatible with the boundary conditions on $\partial M$.  Since (at least when it is allowed by the boundary conditions) empty AdS${}_{d+1}$ is a solution invariant under all symmetries one might expect that the corresponding central extensions are trivial.

This turns out to be true for $d > 2$.  Indeed, any Killing field of AdS${}_{d+1}$ automatically satisfies our definition of an asymptotic symmetry (at least for boundary conditions $\phi^{(0)}=0$ and $\gamma^{(0)}_{ij}$ the metric on the Einstein static universe).  But for $d=2$ there are additional asymptotic Killing fields that are not Killing fields of empty
AdS${}_{3}$.  This is because all $d=2$ boundary metrics $\gamma^{(0)}_{ij}$ take the form $ds^2 = g_{uv} du dv$ when written in terms of null coordinates, making manifest that any vector field $\hat \xi^u = f(u)$, $\hat \xi^v = g(v)$ is a conformal Killing field of $\gamma^{(0)}_{ij}$.  This leads to an infinite-dimensional asymptotic symmetry group, which is clearly much larger than the group $SO(2,2)$ of isometries of AdS${}_3$.

Thus as first noted in \cite{Brown:1986nw} there can be a non-trivial central extension for $d=2$.  In this case, one can show that up to the above-mentioned zero-point shifts all central extensions are parametrized by a single number $c$ called the \subind{central charge}{central extension}.  (When parity symmetry is broken, there can be separate left and right central charges $c_L, c_R$.)   Ref  \cite{Brown:1986nw} calculated this central charge using Hamiltonian methods, but we will follow \cite{Balasubramanian:1999re} and work directly with the boundary stress tensor.

Since the charges $Q[\xi]$ generate (bulk) diffeomorphisms along $\xi$, and since the charges themselves are built from $T_\mathrm{bndy}^{ij}$, the entire effect is captured by computing the action of a bulk diffeomorphism $\xi$ on $T_\mathrm{bndy}^{ij}$.  As noted in section \ref{sec:diffeos}, the action of $\xi$ on boundary quantities generally involves both a diffeomorphism $\hat \xi$ along the boundary and a change of conformal frame.  And as we have seen, for even $d$ changes of conformal frame act non-trivially on $T_\mathrm{bndy}^{ij}$.  For $g_{uv} = -1$ a direct calculation gives
\begin{align}
T_{\mathrm{bndy}\ uu} &\rightarrow T_{\mathrm{bndy}\ uu} + \left( 2T_{\mathrm{bndy}\ uu}  \partial_u \xi^u  + \xi^u \partial_u T_{\mathrm{bndy}\ uu} \right) - \frac{c}{24\pi} \partial_u^3 \xi^u \cr
T_{\mathrm{bndy}\ vv} &\rightarrow T_{\mathrm{bndy}\ vv} + \left( 2T_{\mathrm{bndy}\ vv}  \partial_v \xi^v  + \xi^v \partial_v T_{\mathrm{bndy}\ vv} \right) - \frac{c}{24\pi} \partial_v^3 \xi^v,
\end{align}
where $c=3\ell/2G$.  The term in parenthesis is the tensorial part of the transformation while the final (so called anomalous) term is associated with the \ind{conformal anomaly} $a_{(2)}=-(c/24\pi) {\cal R}$.

It is traditional to Fourier transform the above components of the stress tensor to write the charge algebra as the (double) \ind{Virasoro algebra}
\begin{align}
i\{ L_m, L_n\} &= (m-n)L_{m+n} + \frac{c}{12} m(m^2-1) \delta_{m+n,0}, \\ i\{ \bar L_m, \bar L_n \} &= (m-n)\bar L_{m+n} + \frac{c}{12} m(m^2-1) \delta_{m+n,0},
\end{align}
where $\{L_n, \bar L_m\}~=~0$ and
\begin{equation}
L_n = -\frac{1}{2\pi}\int_{S^1} e^{iun} T_{\mathrm{bndy} \ uu} du, \ \  \ \bar L_n = -\frac{1}{2\pi}\int_{S^1} e^{ivn} T_{\mathrm{bndy} \ vv} dv.
\end{equation}
Here we have take $\partial M = S^1 \times {\mathbb R}$ so that the dynamics requires both $T_{uu}$ and $T_{vv}$ to be periodic functions of their arguments.  We have taken this period to be $2\pi$.

The anomalous transformation of $T_\mathrm{bndy}^{ij}$ leads to interesting zero-points for certain charges.  Suppose for example we take $T_\mathrm{bndy}^{ij}$ to vanish for the Poincar\'e patch of empty AdS${}_3$ in the conformal frame where the boundary metric is (uncompactified) Minkowski space.  Then since $S^1 \times {\mathbb R}$ is (locally) conformal to Minkowski space, we can use the conformal anomaly to calculate $T_\mathrm{bndy}^{ij}$ for empty AdS${}_3$ with Einstein static universe boundary metric.  One finds that the resulting energy does not vanish.  Instead, $E_{{\rm global \ AdS}_3} =  - c/12\ell = - 1/8G$ so that $E=0$ for the so-called $M=0$ Ba\~nados-Teitelboim-Zanelli \ind{(BTZ) black hole} \cite{Banados:1992wn,Banados:1992gq}.  The offset in \eqref{AdSSM} arises from similarly setting $T_\mathrm{bndy}^{ij} =0$ for empty AdS${}_5$ in the conformal frame where the boundary metric is (uncompactified) Minkowski space.

\section{The algebra of boundary observables and the AdS/CFT correspondence}
\label{adscft}

We have shown above how the boundary stress tensor can be used to construct charges $Q[\xi]$ associated with any asymptotic symmetry $\xi$ of a theory of asymptotically locally anti-de Sitter spacetimes.  The $Q[\xi]$ are conserved (perhaps, up to c-number anomaly terms) and generate the asymptotic symmetry $\xi$ under the action of the Peierls bracket (or equivalently, under the Poisson bracket).  Therefore the $Q[\xi]$ are equivalent to the Hamiltonian charges that we could derive using techniques analogous to those described in chapter 17 for asymptotically flat spacetimes.  Conversely, boundary stress tensor methods can also be applied in the asymptotically flat context \cite{Mann:2005yr,Mann:2006bd,Mann:2008ay}.  Readers interested in direct Hamiltonian approaches to AdS charges should consult \cite{Henneaux:1984xu,Henneaux:1985tv,Brown:1986nw}; see also \cite{Abbott:1981ff,Gibbons:1982jg,Gibbons:1983aq,Ashtekar:1984zz,Ashtekar:1999jx,Katz:1996nr,Deruelle:2004mv} for other covariant approaches.

We chose to use boundary stress tensor methods for two closely related reasons.  The first is that, in addition to its role in constructing conserved charges, the local boundary field $T^{ij}_\mathrm{bndy}$ turns out to contain useful information on its own.  For example, it plays a key role in the hydrodynamic description of large AdS black holes known as the \ind{fluid/gravity correspondence} \cite{Bhattacharyya:2008jc} (which may be considered a modern incarnation of the so-called membrane paradigm \cite{Thorne:1986iy}).  The extra information in $T_\mathrm{bndy}^{ij}$  appears at the AdS boundary $\partial M$ due to the fact that all multipole moments of a given field decay near $\partial M$ with the same power law; namely, the one given by the $\gamma^{(d)}$ term in the Fefferman-Graham expansion \eqref{eq:FGexpansion}.  This is in striking contrast with the more familiar situation in asymptotically flat spacetimes where the large $r$ behavior is dominated by the monopole terms, with sub-leading corrections from the dipole and higher order multipoles.  Indeed, while as noted above similar boundary stress tensor techniques can be employed in asymptotically flat spacetimes, the asymptotically flat boundary stress tensor contains far less information.

The second reason is that both $T^{ij}_\mathrm{bndy}$ and $\Phi_\mathrm{bndy}$ play fundamental roles in the \ind{AdS/CFT correspondence} \cite{Maldacena:1997re} (see especially \cite{Witten:1998qj}).  Any treatment of asymptotic AdS charges would be remiss without at least mentioning this connection, and we take the opportunity below to give a brief introduction to AdS/CFT from the gravity side.  This turns out to be straightforward using the machinery described thus far.  Indeed, the general framework requires no further input from either string theory or conformal field theory and should be readily accessible to all readers of this volume.  As usual, we consider bulk gravity coupled to a single bulk scalar and fix both $\gamma_{ij}^{(0)}$ and $\phi^{(0)}$ as boundary conditions. We refer to $\gamma_{ij}^{(0)}$ and $\phi^{(0)}$ as boundary sources below. More general boundary conditions may be thought of as being dual to CFTs with additional interactions \cite{Witten:2001ua} or coupled to additional dynamical fields \cite{Gubser:2002zh,Witten:2003ya,Compere:2008us}, though we will not go into the details here.

The only new concept we require is that of the the algebra ${\cal A}_\mathrm{bndy}$ of \ind{boundary observables}, which is just the algebra generated by $T^{ij}_\mathrm{bndy}$ and $\Phi_\mathrm{bndy}$ under the Peierls bracket.  Here we mean that we consider the smallest algebra containing both
$T^{ij}_\mathrm{bndy}$ and $\Phi_\mathrm{bndy}$ which is closed under finite flows; i.e., under the classical analogue of the quantum operation $e^{iA}Be^{-iA}$.  A key property of ${\cal A}_\mathrm{bndy}$ follows from the fact that the bulk equations of motion are completely independent of the choice of conformal frame $\Omega$.  Thus, up to the usual conformal anomalies, under any change of conformal frame the boundary observables transform only by rescaling with a particular power of $e^{-\sigma}$ known as the \subind{conformal dimension}{conformal transformation} ($d$ for $T_\mathrm{bndy}^{ij}$, and $\Delta_+$ for $\Phi_\mathrm{bndy}$), with the boundary sources transforming similarly with conformal weights zero for $\gamma^{(0)}_{ij}$ and $\Delta_-$ for $\phi^{(0)}$. (In defining the conformal dimension it is conventional not to count the $\pm 2$ powers of $e^{-\sigma}$ associated with the indices on $T_\mathrm{bndy}^{ij}$ and $\gamma^{(0)}_{ij}$.) In this sense the theory of ${\cal A}_\mathrm{bndy}$ is invariant (or, perhaps better, covariant) under all changes of boundary conformal frame. Of course we have already shown that when the boundary observables admit a conformal Killing field $\hat \xi$, the corresponding transformation is generated by the associated $Q[\xi]$ from \eqref{eq:Q}.  Now since the charges $Q[\xi]$ are built from $T_\mathrm{bndy}^{ij}$ and $\Phi_\mathrm{bndy}$ they also lie in the algebra ${\cal A}_\mathrm{bndy}$. When $\hat \xi$ can be chosen to be everywhere timelike, this immediately implies that ${\cal A}_\mathrm{bndy}$ is also closed under time evolution.  This last property can also be shown much more generally; see e.g. \cite{Marolf:2008mf}.

We now extract one final property of the algebra ${\cal A}_\mathrm{bndy}$.
From the expression \eqref{GFs} in terms of Green's functions, it is clear that the Peierls bracket $\{ A,B \}$ of two observables vanishes on any solution $s$ for which $A,B$ are  outside each other's light cones; i.e., when the regions on which $A,B$ are supported cannot be connected by any causal curve.  Furthermore, as shown in \cite{Gao:2000ga} the null energy condition implies that two boundary points $x,y$ can be connected by a causal curve through the bulk only when they can also be connected by a causal curve lying entirely in the boundary.  It follows that the algebra ${\cal A}_\mathrm{bndy}$ satisfies the usual definition of locality for a field theory on $\partial M$; namely that Peierls brackets vanish outside the light cones defined by the boundary metric $\gamma_{ij}^{(0)}$.

Though we have so far worked entirely at the classical level, let us now assume that all of the above properties persist in the quantum theory.  We then have a conformally covariant algebra of operators ${\cal A}_\mathrm{bndy}$ with closed dynamics, local commutation relations on $\partial M$, and a stress tensor $T_\mathrm{bndy}^{ij}$ that generates all conformal symmetries.  In other words, we have a local conformal field theory on $\partial M$.

This is the most basic statement of the AdS/CFT correspondence.  {\it Any} bulk AlAdS quantum gravity theory in which the above classical properties continue to hold defines a conformal field theory (CFT) through its algebra ${\cal A}_\mathrm{bndy}$ of boundary observables.  Now, we should remark that the AdS/CFT correspondence as used in string theory goes one step further.  For certain specific bulk theories it identifies the so-called dual CFT as a particular known theory defined by its own Lagrangian with a definite field content.  For example, when the bulk is type IIB string theory asymptotic to a certain AdS${}_5$ $\times \ S^5$ solution, the corresponding CFT is just ${\cal N} =4$ super-Yang-Mills.  We will not go into further details here, though the interested reader may consult various reviews such as \cite{Aharony:1999ti,D'Hoker:2002aw,Polchinski:2010hw}.

On the other hand, even without having a separate definition of the CFT, the above observations already have dramatic implications for the bulk quantum gravity theory.  In particular, the statement that ${\cal A}_\mathrm{bndy}$ is closed under time evolution runs completely counter to one's usual intuition regarding field theory with a boundary.  We usually think that most of the dynamical degrees of freedom live in the bulk spacetime, with perhaps only a small subset visible on the boundary at any time.  In particular, we expect any signal present on the boundary at time $t_0$ to then propagate into the bulk and (at least for some time) to essentially disappear from the algebra of boundary observables.  Since ${\cal A}_\mathrm{bndy}$ is closed under time evolution, it is clear that this is simply not the case in our quantum gravity theory.  The difference arises precisely from the fact that the gravitational Hamiltonian (and more generally any $Q[\xi]$) is a pure boundary term.  This property was called {\it \ind{boundary unitarity}} in \cite{Marolf:2008mf}.  See also \cite{Marolf:2008mg} for further discussion of this point.

The reader should take care to separate boundary unitarity from the possible claim that ${\cal A}_\mathrm{bndy}$ captures the complete set of bulk observables.  The two ideas are logically separate, as there can in principle be additional bulk observables ${\cal A}_\mathrm{other}$ so long as they do not mix dynamically with those in ${\cal A}_\mathrm{bndy}$.  One says that the possible values of ${\cal A}_\mathrm{other}$ define superselection sectors with respect to ${\cal A}_\mathrm{bndy}$ \cite{Marolf:2008tx}.  But any such additional observables are clearly very special.  The requirement that they not affect ${\cal A}_\mathrm{bndy}$ strongly suggests that at least semi-classically such observables have to do only with properties of spacetime hidden from the boundary behind both past and future horizons \cite{Marolf:2012xe}.  In particular, any degrees of freedom that determine whether black holes are connected by (non-traversable) wormholes seem likely to lie in ${\cal A}_\mathrm{other}$.  On the other hand, in perturbation theory about empty AdS (or even about solutions that are empty AdS in the far past) one may show that ${\cal A}_\mathrm{other}$ is indeed empty \cite{Marolf:2008mf}.

\section*{Acknowledgements}
\edit{We wish to thank Kevin Kuns for spotting minus sign errors in a previous version of this chapter.}  This work was supported in part by the National Science Foundation under Grant Nos PHY11-25915 and PHY08-55415, and by funds from the University of California. DM also thanks the University of Colorado, Boulder, for its hospitality during this work.

\bibliographystyle{JHEP}
\bibliography{Ch21bib}

\end{document}